\documentstyle[manuscript,eqsecnum,aps,version2]{revtex}
\draft
\begin{document}

\preprint{HEP-95-18}
\title
\bf Spherically Symmetric Random Walks II. \\
\bf Dimensionally Dependent Critical Behavior
\endtitle
\author{Carl M. Bender}
\instit
Department of Physics, Washington University, St. Louis, MO 63130
\endinstit
\author{Stefan Boettcher}
\instit
Department of Physics, Brookhaven National Laboratory, Upton, NY 11973
\endinstit
\author{Peter N. Meisinger}
\instit
Department of Physics, Washington University, St. Louis, MO 63130
\endinstit
\medskip
\centerline{\today}
\abstract
A recently developed model of random walks on a $D$-dimensional
hyperspherical lattice, where $D$ is {\sl not} restricted to integer values,
is extended to include the possibility of creating and annihilating random
walkers. Steady-state distributions of random walkers are obtained for all
dimensions $D>0$ by solving a discrete eigenvalue problem. These distributions 
exhibit dimensionally dependent critical behavior as a function of the birth 
rate. This remarkably simple model exhibits a second-order phase transition 
with a nontrivial critical exponent for all dimensions $D>0$.
\endabstract


\section{INTRODUCTION}
\label{s1}
In previous papers \cite{BeBoM,BeCoMe,LET} we analyze a class of models of 
$D$-dimensional spherically symmetric random walks, where $D$ is {\sl not
restricted to integer values.} In this paper we extend these models to allow
for the creation and annihilation of random walkers. We demonstrate that
these extended models exhibit critical behavior as a function of the birth
rate of walkers. The critical coefficients depend on the value of the
dimension $D$. In the next paper in this series we apply these ideas to the
study polymer growth in $D$ dimensions in the vicinity of a hyperspherical
adsorbing boundary.

The random walks in Refs.~\cite{BeBoM,BeCoMe,LET} take place on an infinite set
of regions labeled by the integer $n$, $n=1,~2,~3,~\ldots$. If the random
walker is in $region~n$ at time $t$, then at time $t+1$ the walker must move
outward to $region~n+1$ with probability $P_{\rm out}(n)$ or inward to
$region~n-1$ with probability $P_{\rm in}(n)$, where
\begin{eqnarray}
P_{\rm out}(n)+P_{\rm in}(n)=1
\label{e1.1}
\end{eqnarray}
so that probability is conserved. [We take $P_{\rm out}(1)=1$ and $P_{\rm
in}(1)=0$ to enforce the requirement that a walker in the central region
($n=1$) must move outward at the next step.] Let $C_{n,t;m}$ represent the
probability that a random walker who begins in the $m$th region at $t=0$
will be in the $n$th region at time $t$. The probability $C_{n,t;m}$ then
satisfies the difference equation
\begin{eqnarray}
C_{n,t;m}=\cases{P_{\rm in}(n+1) C_{n+1,t-1;m}+P_{\rm
out}(n-1)C_{n-1,t-1;m}& $(n\geq 2),$\cr 
\noalign{\medskip}
P_{\rm in}(2) C_{2,t-1;m}& $(n=1)$\cr}
\label{e1.2}
\end{eqnarray}
and the initial condition
\begin{eqnarray}
C_{n,0;m}=\delta_{n,m}.
\label{e1.3}
\end{eqnarray}

To formulate a model of spherically symmetric random walks in $D$-dimensional 
space we take {\sl region} $n$ to be the volume bounded by two
concentric $D$-dimensional hyperspherical surfaces of radii $R_{n-1}$ and
$R_n$. In Ref.~\cite{BeBoM} we take the probabilities of moving out or in to
be in proportion to the hyperspherical surface areas bounding $region~n$.
Let $S_D(R)$ represent the surface area of a $D$-dimensional hypersphere
\begin{eqnarray}
S_D(R)={{2\pi^{D/2}}\over{\Gamma(D/2)}} R^{D-1}.
\nonumber
\end{eqnarray}
Then, for $n>1$,
\begin{eqnarray}
P_{\rm out}(n)={{S_D(R_n)}\over{S_D(R_n)+S_D(R_{n-1})}}=
{{R_n^{D-1}}\over{R_n^{D-1}+R_{n-1}^{D-1}}}
\label{e1.5}
\end{eqnarray}
and
\begin{eqnarray}
P_{\rm in}(n)={{S_D(R_{n-1})}\over{S_D(R_n)+S_D(R_{n-1})}}=
{{R_{n-1}^{D-1}}\over{R_n^{D-1}+R_{n-1}^{D-1}}}.
\label{e1.6}
\end{eqnarray}
As we discussed earlier, for the special case $n=1$ we define
\begin{eqnarray}
P_{\rm out}(1)=1\quad{\rm and}\quad P_{\rm in}(1)=0.
\label{e1.7}
\end{eqnarray}
The choices in Eqs.~(\ref{e1.5})-(\ref{e1.7}) satisfy the requirement that
probability be conserved because they obey Eq.~(\ref{e1.1}).

{}For dimensions other than $D=1$ and $D=2$, when we substitute 
Eqs.~(\ref{e1.5})-(\ref{e1.7}) into Eq.~(\ref{e1.2}) and take $R_n=n$, we
obtain a difference equation that cannot be solved in closed form. Thus, in
Ref.~\cite{BeCoMe} we proposed that the probabilities in
Eqs.~(\ref{e1.5})-(\ref{e1.7}) be replaced by bilinear functions of $n$,
which are are uniformly good approximations to $P_{\rm out}(n)$ and $P_{\rm
in}(n)$ in the range $D>0$ when $R_n=n$:
\begin{eqnarray}
P_{\rm out}(n)={n+D-2\over 2n+D-3}\quad {\rm and}\quad
P_{\rm in}(n)={n-1\over 2n+D-3}.
\label{e1.8}
\end{eqnarray}
Now, the difference equation initial-value problem (\ref{e1.2}) and
(\ref{e1.3}) for the probabilities $C_{n,t;m}$ can be solved in closed form:
\begin{eqnarray}
C_{n,t;m} = {(2n+D-3)\Gamma^2(D-1)\Gamma(m)\over 2^{D-1}\Gamma^2(D/2)
\Gamma(m+D-2)}\int_{-1}^1 dx\; (1-x^2)^{(D-2)/2}x^t
{\cal C}_{n-1}^{( {D-1\over 2})}(x) {\cal C}_{m-1}^{( {D-1\over 2})}(x),
\label{e1.9}
\end{eqnarray}
where ${\cal C}_n^{(\alpha)}(x)$ is a Gegenbauer polynomial \cite{A+S}. From 
the solution in Eq.~(\ref{e1.9}) one can obtain closed-form expressions for
spatial and temporal moments of the random walk \cite{BeCoMe}.

In this paper we generalize the difference equation (\ref{e1.2}) to include
the possibility of creation and annihilation of random walkers. We allow
random walkers to give birth in $region~1$ with birth rate $a$ and to die in
all other regions with uniform death rate $z$. Birth rates and death rates
are properties of populations rather than of single individuals. Thus,
rather than solving  Eq.~(\ref{e1.2}) as an initial-value problem for a {\sl
single} random walker who starts in {\sl region} $m$, we are going to study
a large population of random walkers, all of whom obey this difference
equation. We represent this population of random walkers by a distribution
$G_{n,t}$, which denotes the number of random walkers in {\sl region} $n$ at
time $t$. The distribution $G_{n,t}$ satisfies the same recursion relation
as $C_{n,t;m}$ except for the factors of $a$ and $z$:
\begin{eqnarray}
G_{n,t}=\cases{zP_{\rm in}(n+1) G_{n+1,t-1}+zP_{\rm out}(n-1)G_{n-1,t-1}&
$(n\geq 3),$\cr 
\noalign{\medskip}
zP_{\rm in}(3)G_{3,t-1}+aG_{1,t-1}&
$(n=2),$\cr 
\noalign{\medskip}
zP_{\rm in}(2) G_{2,t-1}& $(n=1),$\cr}
\label{e1.10}
\end{eqnarray}
where we have set $P_{\rm out}(1)=1$. Note that the function $G_{n,t}$ must
be positive for all $n$ and $t$. Aside from the requirement that $G_{n,0}$,
the initial distribution of random walkers, be normalizable it is arbitrary.
We are not concerned with the detailed structure of the initial
distribution; rather, we are interested in the asymptotic behavior of
distributions as $t\to\infty$. The specific choice of the initial
distribution is unimportant because a random walk is a diffusive
(dissipative) process and details of $G_{n,0}$ are irretrievably lost as
time evolves; all initial distributions lead to the same large-time
behavior. This behavior is determined by the details of the random walk
process itself.

In this model random walkers are created or destroyed at a given site {\sl
in proportion to} the number of walkers at that site, where $a$ and $z$ are
the constants of proportionality. Technically speaking, $a$ acts as a birth
rate if $a>1$; if $a<1$, it is really a death rate. A similar interpretation
applies to $z$. We are particularly interested in steady-state solutions of
Eq.~(\ref{e1.10}); the existence of such solutions imposes a relationship
between the birth rate and the death rate.

In Sec.~\ref{s2} we perform numerical and analytical studies of
Eq.~(\ref{e1.10}) for arbitrary choice of $P_{\rm out}(n)$ and $P_{\rm
out}(n)$. We study two quantities: $N_t$, the total number of random walkers
at time $t$, and $F_t$, the fraction of random walkers in $region~1$ at time
$t$. We show that the (physically relevant) positive quadrant of the $(a,z)$
plane is partitioned by two intersecting critical curves into four regions
(or phases) characterized by the behavior of $N_t$ and $F_t$ as
$t\to\infty$. (Some detailed asymptotic studies of the large-$t$ behavior
are given in the Appendix.) The intersection of these two critical curves is
the critical point $(a_{\rm c},z_{\rm c})$ for steady-state distributions of
random walkers. Although the location of this critical point is a function
of the dimension $D$, the qualitative features of this phase diagram are
generic.

We obtain a solvable model of random walks for arbitrary dimension $D$ by
using the uniform approximations for $P_{\rm out}(n)$ and $P_{\rm in}(n)$
proposed in Ref.~\cite{BeCoMe} and given in Eq.~(\ref{e1.8}). In
Sec.~\ref{s3} we analyze  this model for the simple case $D=1$. We
analytically determine the features of the $(a,z)$ phase diagram and verify
that random walk distributions exhibit critical behavior. Then, in
Sec.~\ref{s4} we use generating-function methods to solve Eq.~(\ref{e1.10})
for arbitrary $D$. We cannot perform a global analysis of the $(a,z)$ phase
diagram as in the case $D=1$ but we can perform a local analysis in the
vicinity of the critical point. From this local analysis we show that a
second-order phase transition occurs at the critical point.

Specifically, we show that near the critical birth rate $a_{\rm c}$, the
steady state distribution fraction $F(a)=\lim_{t\to\infty}F_t$ behaves like
\begin{eqnarray}
{}F(a)\sim C(D)(a-a_{\rm c})^{\nu}\quad (a\to a_{\rm c}^+,~D\neq 2,4),
\label{e1.11}
\end{eqnarray}
where the multiplicative constant ${\rm C}(D)$ depends on the dimension $D$.
The critical exponent $\nu$ also depends on the dimension $D$:
\begin{eqnarray}
\nu=\cases{{D\over 2-D}&$(0<D<2),$\cr
\noalign{\medskip}
{2\over D-2}&$(2<D<4),$\cr 
\noalign{\medskip}
1 &$(D>4).$\cr} \label{e1.12}
\end{eqnarray}
There is no critical exponent for the special cases $D=2,4$; instead, we
find that as $a\to a_{\rm c}^+$
\begin{eqnarray}
{}F(a)\sim\cases{\displaystyle
{{\rm constant}\over a-a_{\rm c}} e^{-{2\over a-a_{\rm c}}}&$(D=2),$\cr
\noalign{\medskip}
\displaystyle {a-a_{\rm c} \over 12\log \left({1\over a-a_{\rm c}}\right )} 
&$(D=4).$\cr}
\label{e1.13}
\end{eqnarray}

In general, formulating simplified $D$-dimensional statistical models is
useful for understanding aspects of critical phenomena exhibited actual
physical systems. Indeed, any solvable statistical model that exhibits
nontrivial critical behavior is worthy of study \cite{Bax}. In the next
paper in this series we apply the results of this paper to the study of
polymer growth in $D$ dimensions. There we extend to arbitrary dimension the
earlier results for $D=1$ \cite{Priv} and $D=2$ \cite{BoMo,Bo}.

\section{RANDOM WALKS WITH BIRTH AND DEATH}
\label{s2}

In this section we discuss the general properties of the spherically
symmetric random walk model defined by Eq.~(\ref{e1.10}) in which random
walkers may be created and annihilated. Let 
\begin{eqnarray}
N_t=\sum_{n=1}^{\infty}G_{n,t}
\label{e2.1}
\end{eqnarray}
be the total number of random walkers at time $t$. We restrict our attention
to initial distributions for which $N_0$ is finite so that $N_t$ is finite
for all $t$. Let 
\begin{eqnarray}
{}F_t=G_{1,t}/N_t
\label{e2.2}
\end{eqnarray}
represent the {\sl fraction} of all random walkers in $region~1$ at time
$t$.

Numerical \cite{Num} and analytical studies of the quantities $N_t$ and
$F_t$ as $t\to\infty$ reveal that, independent of the initial distribution
of random walkers, the asymptotic behaviors of $N_t$ and $F_t$ are
determined by the values of $a$ and $z$. Specifically, we obtain the generic
result that for any value of $D$ the positive quadrant of the $(a,z)$ plane
is partitioned into four distinct regions by two boundary curves as shown in
{}Fig.~\ref{f1}.

One of the boundary curves, which we have labeled $B_1$ in Fig.~\ref{f1}, is
a straight line passing through the origin. To the left of $B_1$ we find
that $F_t$ vanishes as $t\to\infty$; to the right of $B_1$ we find that
$F_t$ approaches a positive finite value as $t\to\infty$. For $D\leq 2$ the
equation for the boundary line $B_1$ is $z=a$; as $D$ increases beyond $2$
the boundary line remains straight but the slope of $B_1$ begins to decrease
with increasing $D$. As we will see, the transition that occurs at $D=2$ is
a reflection of Polya's theorem \cite{Polya}, which states that when $D>2$ the
probability of an individual random walker visiting $region~1$ more than
once is less than unity.

The second boundary curve shown in Fig.~\ref{f1} is labeled $B_2$. This
curve consists of two parts: The first part is a straight-line segment,
$z=1$, extending from the $z$ axis to the boundary line $B_1$. This line
segment connects to the second part, which is a curve that approaches $z=0$
as $a\to\infty$. The equation describing the second part depends on $D$.
[For $D=1$ this curve is given by $z=2a/(a^2+1)~(a\geq 1)$, as we will show
in Sec.~\ref{s3}.] Above the boundary $B_2$ we find that $N_t\to\infty$ as
$t\to\infty$; below $B_2$ we find that $N_t\to 0$ as $t\to\infty$. On $B_2$
the total number of walkers approaches a finite distribution $N(a)$ as
$t\to\infty$. On the curved portion of $B_2$ the function $N(a)$ is
positive; on the straight-line portion of $B_2$ the function $N(a)$ is
positive for $D>2$ while $N(a)=0$ for $D\leq 2$. This transition at $D=2$ is
yet another manifestation of Polya's theorem.

While detailed studies of the large-$t$ asymptotic behavior of $N_t$ and
$F_t$ are given in the Appendix, many of the qualitative features of
{}Fig.~\ref{f1} can be derived directly from an analysis of Eq.~(\ref{e1.10}).
To determine the boundary line $B_1$ we introduce a change of variable in
Eq.~(\ref{e1.10}): \begin{eqnarray}
G_{n,t}=z^tH_{n,t}.
\label{e2.3}
\end{eqnarray}
The distribution $H_{n,t}$ satisfies the recursion relation for a $D$-dimensional 
spherically symmetric random walk with a birth rate $a/z$ in
$region~1$ and no births or deaths occurring in any other region:
\begin{eqnarray}
H_{n,t}=\cases{P_{\rm in}(n+1) H_{n+1,t-1}+P_{\rm out}(n-1)H_{n-1,t-1}&
$(n\geq 3),$\cr 
\noalign{\medskip}
P_{\rm in}(3)H_{3,t-1}+{a\over z}H_{1,t-1}& $(n=2),$\cr 
\noalign{\medskip}
P_{\rm in}(2) H_{2,t-1}& $(n=1).$\cr}
\label{e2.4}
\end{eqnarray}

Let $\Pi_1(D)$ denote the probability that a random walker in $region~1$
will eventually return to $region~1$. Suppose a random walk satisfying
Eq.~(\ref{e2.4}) begins at $t=0$. Of the $H_{1,0}$ walkers who begin in
$region~1$, only a fraction $\Pi_1(D)$ of them will eventually return to
$region~1$ to give birth to new walkers at the rate $a/z$. Of these new
walkers, again only a fraction $\Pi_1(D)$ of them will return to $region~1$
to give birth again, and so on. Hence, to find the total number of random
walkers who are ever born we must sum a geometric series whose geometric
ratio is the quantity $a\Pi_1(D)/z$. If this quantity is less than $1$, the
geometric series converges and the total number of random walkers ever born
is finite. As time $t$ increases, the random walkers diffuse away from
$region~1$. Thus, the ratio
\begin{eqnarray}
{}F_t={G_{1,t}\over\sum_{n=1}^{\infty}G_{n,t}}={H_{1,t}\over\sum_{n=1}^{\infty} H_{n,t}}
\nonumber
\end{eqnarray}
vanishes as $t\to\infty$. On the other hand, if the quantity $a\Pi_1(D)/z$
is greater than $1$, both $H_{1,t}$ and $\sum_{n=0}^{\infty} H_{n,t}$
diverge at the same rate and the ratio $F_t$ approaches a nonzero limit
(that lies between 0 and 1) as $t\to\infty$.

The transition between $F_t\to 0$ and $F_t\to finite~limit$ occurs on the line
\begin{eqnarray}
z=a\Pi_1(D).
\label{e2.6}
\end{eqnarray}
This is the equation of the boundary line $B_1$. Polya's theorem states that
for any random walk $\Pi_1(D)=1$ when $D\leq 2$ and $\Pi_1(D)<1$ when $D>2$.
This theorem explains the transition in the slope of the line $B_1$ at
$D=2$. In the spherically symmetric random walk model discussed in
Ref.~\cite{BeBoM}, where $P_{\rm out}(n)$ and $P_{\rm in}(n)$ are given in
Eqs.~(\ref{e1.5}-\ref{e1.7}), it was shown that
\begin{eqnarray}
\Pi_1(D)=1-1/\zeta(D-1)\qquad (D\geq 2)
\label{e2.6a}
\end{eqnarray}
($\zeta$ is the Riemann Zeta function); in the random walk model discussed
in Ref.~\cite{BeCoMe}, where $P_{\rm out}(n)$ and $P_{\rm in}(n)$ are given
in Eq.~(\ref{e1.8}), it was shown that
\begin{eqnarray}
\Pi_1(D)=1/(D-1) \qquad (D\geq 2).
\label{e2.6b}
\end{eqnarray}
Numerical computation confirms the slope of the boundary line $B_1$ for both
models (see Figs.~\ref{f2}-\ref{f9}).

The shape of the curved part of the boundary $B_2$ in Fig.~\ref{f1} depends
on the dimension $D$ and on the choice of the functions $P_{\rm out}(n)$ and
$P_{\rm in}(n)$. It is not universal. However, the straight-line portion of
the boundary $B_2$ is universal and is easy to understand for any $D$.
Points $(a,z)$ such $a<a_{\rm c}$ and $z$ near $1$ lie to the left of $B_1$.
Thus, $F_t$, the fraction of random walkers in $region~1$, becomes
vanishingly small as $t\to\infty$. Hence, the effect of the birth rate $a$
on the total number of walkers is negligible. The growth or decay of the
total number of walkers only depends on the magnitude of $z$; if $z<1$ then
$N_t\to 0$ as $t\to\infty$, and if $z>1$ the $N_t\to\infty$ as $t\to\infty$.

On the straight-line portion of the curve $B_2$, where $a<a_{\rm c}$ and
$z=1$, the limiting value of $N_t$ depends on the dimension $D$. If $D\leq
2$ then $a_{\rm c}=1$. Thus, on this portion of $B_2$ a fraction $1-a$ of
random walkers who arrive in $region~1$ at a given time step must die at the
next time step. But by Polya's theorem {\sl all} random walkers visit
$region~1$ repeatedly. Hence, the total number of random walkers $N_t$ must
vanish as $t\to\infty$. On the other hand, if $D>2$ we have $\Pi_1(D)<1$.
Thus, the fraction $1-\Pi_1(D)$ of random walkers who originate in $region~1$
{\rm never return} to $region~1$. Thus, these random walkers never die
because $z=1$. Hence, $N_t$ approaches a finite positive number as
$t\to\infty$.

We find numerically that as we cross the boundary line $B_1$, the limiting
value of the function $F_t$ as $t\to\infty$ is continuous. We are
particularly interested in crossing from one side of $B_1$ to the other
along the boundary curve $B_2$ that divides the upper region, where
$N_t\to\infty$, and the lower region, where $N_t\to 0$ as $t\to\infty$. We
focus on this curve $B_2$ because it is only on this curve that a steady
state is reached as $t\to\infty$. Along this boundary curve the limiting
value of $F_t$ undergoes a second-order phase transition at the critical
point $(a_{\rm c},z_{\rm c})$, which is situated at the intersection of
$B_1$ and $B_2$. On the curve $B_2$ when $a<a_{\rm c}$ the limiting value of
$F_t$ is $0$ (even though the limiting value of $N_t$ may be $0$), and when
$a>a_{\rm c}$ the limiting values of both $N_t$ and $F_t$ on the boundary
curve $B_2$ are {\sl finite} positive numbers. The curved portion of $B_2$
is in fact the locus of all points in the positive quadrant of the $(a,z)$
plane for which the limiting values of both $N_t$ and $F_t$ as $t\to\infty$
are finite and nonzero.

The interpretation of $\lim_{t\to\infty}N_t$ being finite and nonzero is that 
the distribution $G_{n,t}$ approaches a steady-state. In such a steady-state 
there is a balance between random walkers being created in $region~1$
and annihilated in all other regions. This steady-state solution can be
obtained by solving a discrete eigenvalue problem.

Steady-state distributions are special cases of shape-independent
distributions; that is, distributions that do not change shape as they
evolve in time. For such distributions $G_{n,t}/G_{m,t}$ is independent of
$t$ for all $n$ and $m$ so that the relative number of walkers in {\sl
region} $n$ is a time-independent fraction of the total number of walkers.
The time dependence of such distributions is very simple:
\begin{eqnarray}
G_{n,t}=g_n\lambda^t.
\label{e2.7}
\end{eqnarray}
The distribution $g_n$ satisfies the discrete eigenvalue problem
\begin{eqnarray}
\lambda g_n=\cases{P_{\rm in}(n+1)zg_{n+1}+P_{\rm out}(n-1)zg_{n-1}& $(n\geq
3),$ \cr
\noalign{\medskip}
P_{\rm in}(3)zg_3+ag_1& $(n=2),$\cr 
\noalign{\medskip}
P_{\rm in}(2)zg_2& $(n=1),$\cr}
\label{e2.8}
\end{eqnarray}
which is obtained by substituting $G_{n,t}$ in Eq.~(\ref{e2.7}) into 
Eq.~(\ref{e1.10}). Here, the eigenvalue $\lambda$ represents the
multiplicative growth or decay of the total number of walkers that occurs at
each time step. Since we are interested in distributions of random walkers
for which the birth rate balances the death rate (so that the total number
of walkers is constant in time), we must set $\lambda=1$ in
Eq.~(\ref{e2.8}). We solve this eigenvalue equation for the case $D=1$ in
Sec.~\ref{s3} and for the case of arbitrary $D$ in Sec.~\ref{s4}.

\section{ONE-DIMENSIONAL RANDOM WALKS WITH BIRTH AND DEATH}
\label{s3}

In this section we consider the one-dimensional $(D=1)$ version of the discrete 
eigenvalue problem Eq.~(\ref{e2.8}). When $D=1$, Eqs.~(\ref{e1.5}-\ref{e1.7}) 
and (\ref{e1.8}) both reduce to
\begin{eqnarray}
P_{\rm out}(n)=\cases{{1\over 2}& $(n\geq 2),$\cr 
\noalign{\medskip}
1& $(n=1)$\cr} 
\nonumber
\end{eqnarray}
and
\begin{eqnarray}
P_{\rm in}(n)=\cases{{1\over 2}& $(n\geq 2),$\cr 
\noalign{\medskip}
0& $(n=1).$\cr} 
\nonumber
\end{eqnarray}
{}For this case the steady-state distribution obtained by setting $\lambda=1$
in Eq.(\ref{e2.8}) satisfies
\begin{eqnarray}
g_n=\cases{{1\over 2}z g_{n+1}+{1\over 2}zg_{n-1}& $(n\geq 3),$\cr 
\noalign{\medskip}
{1\over 2}zg_3+ag_1& $(n=2),$\cr 
\noalign{\medskip}
{1\over 2}z g_2& $(n=1).$\cr} 
\label{e3.3}
\end{eqnarray}

It is easy to solve the difference equation (\ref{e3.3}) because it is a
linear constant-coefficient equation. Its general solution has the form
\begin{eqnarray}
g_n=A r_-^{n-2}+Br_+^{n-2}\quad (n\geq 2),
\label{e3.4}
\end{eqnarray}
where
\begin{eqnarray}
r_{\pm}^2 - {2\over z} r_{\pm} +1 = 0
\label{e3.5}
\end{eqnarray}
and $A$ and $B$ are arbitrary constants. The solutions to the quadratic
equation (\ref{e3.5}) are
\begin{eqnarray}
r_{\pm} = {1\over z} \left ( 1 \pm \sqrt{1-z^2} \right ) .
\label{e3.6}
\end{eqnarray}
Observe that
\begin{eqnarray}
r_- r_+ = 1.
\label{e3.7}
\end{eqnarray}

Since the total number of random walkers is finite, the sum
$\sum_{n=1}^{\infty}g_n$ exists. From the existence of this sum and
Eq.~(\ref{e3.7}) we may conclude that $r_{\pm}$ are real; if $r_{\pm}$ were
complex then, since they are complex conjugates, we would have $|r_{\pm}|=1$
and the sum would diverge. Furthermore, since $r_+>1$, it follows that
$B=0$.

If we substitute the solution (\ref{e3.4}) with $B=0$ into the special cases
($n=1$ and $n=2$) of Eq.~(\ref{e3.3}), we obtain a relationship between the
birth rate $a$ and the death rate $z$:
\begin{eqnarray}
a=r_+={1\over z}\left ( 1+\sqrt{1-z^2}\right ),\quad {\rm or}\quad
z={2a\over a^2+1}.
\label{e3.8}
\end{eqnarray}
This is the equation for the curved part of $B_2$, the boundary curve
between the upper region where $N_t\to\infty$ and the lower region where
$N_t\to 0$ when $a\geq 1$.

As a function of the birth rate $a$, the fraction $F(a)$ of random walkers
in $region~1$ for the steady-state distribution $g_n$ is given by
\begin{eqnarray}
{}F(a)={g_1\over \sum_{n=1}^{\infty} g_n}={a-1\over a(a+1)}.
\label{e3.9}
\end{eqnarray}
(Note that the overall multiplicative constant $A$ drops out from this
result and is unimportant.) Equation (\ref{e3.9}) is only valid for $a>1$;
if $a\leq 1$ then no nontrivial steady-state solution exists; the limiting
value of $F_t$ as $t\to\infty$ is $0$. Indeed, it is shown in the Appendix
that as $t\to\infty$ the fraction $F_t$ vanishes like $1/\sqrt{t}$ along the
line $B_1$ and like $1/t$ everywhere to the left of $B_1$.

We observe a second-order phase transition in $F(a)=\lim_{t\to\infty}F_t(a)$
as a function of the birth rate $a$; below the critical birth rate $a_{\rm
c}=1$ this fraction vanishes and just above the critical point the fraction
rises linearly with slope $1/2$:
\begin{eqnarray}
{}F(a)\sim{1\over 2}(a-a_{\rm c})\quad (a\to 1^+).
\label{e3.10}
\end{eqnarray}
Hence, at $D=1$ the critical exponent $\nu$ in Eq.~(\ref{e1.11}) is 1 and
the constant $C(1)={1\over 2}$.

\section{$D$-DIMENSIONAL RANDOM WALKS WITH BIRTH AND DEATH}
\label{s4}
In this section we generalize the analysis of the previous section to
arbitrary dimension $D$. When $D\neq 1$ the difference equation (\ref{e2.8})
is no longer a constant-coefficient difference equation and it cannot be
solved in closed form. Hence, we use the method of generating functions to
study steady-state ($\lambda =1$) solutions of this difference equation.

We seek a solution to the $D$-dimensional generalization of Eq.~(\ref{e3.3})
\begin{eqnarray}
g_n=\cases{{n\over 2n+D-1}z g_{n+1}+{n+D-3\over 2n+D-5}zg_{n-1}& $(n\geq 3),$\cr
\noalign{\medskip}
{2\over D+3}zg_3+ag_1& $(n=2),$\cr
\noalign{\medskip}
{1\over D+1}z g_2& $(n=1),$\cr}
\label{e4.1}
\end{eqnarray}
which is obtained by substituting the uniform approximations to $P_{\rm
out}(n)$ and $P_{\rm in}(n)$ given in Eq.~(\ref{e1.8}) into Eq.~(\ref{e2.8})
and setting $\lambda=1$.

{}For a steady-state solution having a finite number of random walkers the
sum $\sum_{n=1}^{\infty} g_n$ exists. We may therefore sum both sides of
Eq.~(\ref{e4.1}) from $n=1$ to $\infty$ and simplify the result:
\begin{eqnarray}
\sum_{n=1}^{\infty} g_n= (a-z)g_1 + z\sum_{n=1}^{\infty} g_n. \label{e4.2}
\end{eqnarray}
Assuming that the sum $\sum_{n=1}^{\infty} g_n$ is nonzero we may
immediately conclude that
\begin{eqnarray}
{}F(a)={g_1\over \sum_{n=1}^{\infty} g_n}= {1-z(a)\over a-z(a)}. \label{e4.3}
\end{eqnarray}
Note that the result in Eq.~(\ref{e4.3}) is valid on the curved part of
$B_2$, where the sum exists and is nonzero; it is also valid on the
straight-line portion of $B_2$ when $D>2$. On the curve $B_2$ we must treat
$z$ as a function of $a$. We emphasize this dependence by writing $z(a)$ and
by treating the fraction $F$ as a function of $a$ only.

To obtain the dependence of $z$ on $a$ along $B_2$ we use generating
function methods. To begin, we simplify Eq.~(\ref{e4.1}) by setting
\begin{eqnarray}
g_n= (2n+D-3) h_n.
\label{e4.4}
\end{eqnarray}
Substituting Eq.~(\ref{e4.4}) into Eq.~(\ref{e4.1}) we obtain
\begin{eqnarray}
(2n+D-3)h_n=\cases{ nz h_{n+1}+(n+D-3)zh_{n-1}& $(n\geq 3),$ \cr 
\noalign{\medskip}
2zh_3+(D-1)ah_1& $(n=2),$\cr 
\noalign{\medskip}
z h_2& $(n=1).$\cr}
\label{e4.5}
\end{eqnarray}
Next, we define a generating function:
\begin{eqnarray}
H(x)=\sum_{n=0}^{\infty} x^n h_{n+1}.
\label{e4.6}
\end{eqnarray}
Note that
\begin{eqnarray}
G(x)= \sum_{n=0}^{\infty} x^n g_{n+1}= \left ( 2x{d\over dx}+D-1\right )
H(x). \label{e4.7}
\end{eqnarray}

Multiplying Eq.~(\ref{e4.5}) by $x^{n-1}$ and summing both sides from $n=3$
to $\infty$, we obtain a first-order inhomogeneous linear differential
equation for $H(x)$:
\begin{eqnarray}
(zx^2-2x+z)H'(x) + (D-1)(zx-1)H(x) = (z-a)(D-1)x h_1.
\label{e4.8}
\end{eqnarray}
To solve Eq.~(\ref{e4.8}) we multiply both sides by the integrating factor
$(x^2z-2x+z)^{D-3\over 2}$. The differential equation then simplifies to
\begin{eqnarray}
{d \over dx}\left [(zx^2-2x+z)^{D-1\over 2}H(x)\right ]=(z-a) g_1 x
(x^2z-2x+z)^{D-3\over 2}.
\label{e4.9}
\end{eqnarray}
The general solution to Eq.~(\ref{e4.9}) is
\begin{eqnarray}
H(x) = (zx^2-2x+z)^{1-D\over 2} \left [ C + (z-a) g_1 \int_0^x ds\, s
(s^2z-2s+z)^{D-3\over 2} \right ],
\label{e4.10}
\end{eqnarray}
where $C$ is an arbitrary constant.

To determine the constant $C$ we observe that $H(0) = h_1 = {g_1\over D-1}$,
from which it follows that
\begin{eqnarray}
C={g_1\over D-1} z^{D-1\over 2}.
\nonumber
\end{eqnarray}
Thus,
\begin{eqnarray}
H(x) = g_1 \left ( x^2-2{x\over z}+1 \right )^{1-D\over 2}
\left [ {1\over D-1} + \left ( 1 - {a\over z}\right ) \int_0^x ds\, s \left
( s^2-2{s\over z}+1\right )^{D-3\over 2} \right ].
\label{e4.12}
\end{eqnarray}
{}Finally, we use Eq.~(\ref{e4.7}) to obtain the generating function $G(x)$:
\begin{eqnarray}
G(x) &=& g_1 \Bigg \{ \left ( x^2-2{x\over z}+1 \right )^{-{D+1\over 2}} (1-
x^2) \left [ 1 + (D-1) \left ( 1 - {a\over z}\right ) \int_0^x ds\, s \left
( s^2-2{s\over z}+1\right )^{D-3\over 2} \right ] \nonumber\\ &&\qquad +{
2x^2\left ( 1-{a\over z}\right )\over x^2-2 {x \over z}+1 }\Bigg \}.
\label{e4.13}
\end{eqnarray}
Assuming that $G(x)$ exists for all $0\leq x\leq 1$, we formally recover
Eq.~(\ref{e4.3}) when we set $x=1$.

Recall the quantities $r_\pm$ defined in Eq.~(\ref{e3.6}) and rewrite
Eq.~(\ref{e4.13}) as
\begin{eqnarray}
G(x) &=& g_1 \Bigg \{ \left [ (r_+ -x)(r_- -x) \right ]^{-{D+1\over 2}} (1-
x^2) \Bigg [ 1 + (D-1) \left ( 1 - {a\over z}\right )\nonumber\\
&&\qquad\times\int_0^x ds\, s( r_+ -s)^{D-3\over 2} (r_- -s)^{D-3\over
2}\Bigg ] +{ 2x^2\left ( 1-{a\over z}\right ) \over ( r_+ -x)(r_- -x)} \Bigg
\}. \label{e4.14}
\end{eqnarray}
{}For $z<1$ the generating function $G(x)$ may be singulare at $x=r_-<1$, in
which case the representation of $G(1)$ as a series will not exist. To
preclude the possibility of such a singularity it is necessary and
sufficient to impose the following eigenvalue condition:
\begin{eqnarray}
1 + (D-1) \left ( 1 - {a\over z}\right ) \int_0^{r_-} ds\, s
( r_+ -s)^{D-3\over 2} (r_- -s)^{D-3\over 2}  =0.
\label{e4.15}
\end{eqnarray}
This condition is clearly necessary. We can verify that it is sufficient by
showing that $G(r_- - \epsilon)$ exists in the limit as $\epsilon\to 0^+$.
To leading order in $\epsilon$ the eigenvalue condition in Eq.~(\ref{e4.15})
becomes
\begin{eqnarray}
1 &+& (D-1) \left ( 1 - {a\over z}\right ) \int_0^{r_- -\epsilon} ds\, s (
r_+ -s)^{D-3\over 2} (r_- -s)^{D-3\over 2} \nonumber\\
&&\qquad\sim -2\left (1-{a\over z}\right )r_- (r_+ - r_-)^{D-3\over 2}
\epsilon^{D-1\over 2}\quad (\epsilon\to 0^+).
\label{e4.16}
\end{eqnarray}
Substituting this asymptotic result into Eq.~(\ref{e4.14}) we see that the
last term, which is of order $\epsilon^{-1}$, exactly cancels.

The eigenvalue condition in Eq.~(\ref{e4.15}) expresses the relation between
$a$ and $z$ that we seek. We can rewrite this condition more compactly by
rescaling the integration variable. Let $s=r_- u$, so that
\begin{eqnarray}
(D-1) r_-^2 \left ( {a\over z} - 1\right ) \int_0^1 du\, u
( 1-u)^{D-3\over 2} (1- r_-^2 u)^{D-3\over 2} =1.
\label{e4.17}
\end{eqnarray}
This integral converges only if $D>1$. We can analytically continue to
values $0<D\leq 1$ [recall that the region of validity of the uniform
approximation in Eq.~(\ref{e1.8}) is $D>0$] by recognizing that this
expression contains the standard integral representation for a
hypergeometric function\cite{A+S}: \begin{eqnarray}
{4~r_-^2 \over D+1 } \left ( {a\over z} - 1\right )
{}_2F_1\left ( {3-D\over 2},2;{D+3\over 2}; r_-^2 \right ) =1. \label{e4.18}
\end{eqnarray}
Using the transformation formulas (especially 15.3.26, 15.2.18, and 15.2.20
in Ref.~\cite{A+S}) for hypergeometric functions, this eigenvalue condition 
can be simplified:
\begin{equation}
1=\left(1-{z\over a}\right) {}_2F_1\left({1\over 2},1;{D+1\over
2};z^2\right). \label{e4.185}
\end{equation}
This implicit equation determines the curve $B_2$ in the $(a,z)$ plane.
However, such a higher transcendental equation cannot be solved for $z$ as
a function of $a$ in closed form. Thus, we perform an asymptotic analysis of
this condition for $z$ near $1$. [As in the previous section, we find that
for $z\to 1^-$ along $B_2$ there is a transition at $z=1$ from nontrivial
steady-state solutions to trivial solutions of the walk equation
(\ref{e4.1}).]

To perform this analysis we let $z=1-\eta$. We then use the following
formula for the analytic continuation of a hypergeometric function:
\begin{eqnarray}
{}_2F_1(a,b;c;\zeta) &=& {\Gamma(c) \Gamma(c-a-b)\over \Gamma(c-a)\Gamma(c-
b)} {}_2F_1(a,b;a+b-c+1;1-\zeta)\nonumber\\
&&~+(1-\zeta)^{c-a-b} {\Gamma(c) \Gamma(a+b-c)\over \Gamma(a)\Gamma(b)}
{}_2F_1(c-a,c-b;c-a-b+1;1-\zeta).
\label{e4.19}
\end{eqnarray}
Next, we substitute the first few terms in the Taylor series of a
hypergeometric function
\begin{eqnarray}
{}_2F_1(a,b;c;\zeta) = {\Gamma(c) \over \Gamma(a)\Gamma(b)}
\sum_{n=0}^{\infty} {\Gamma(a+n)\Gamma(b+n)\over n!\Gamma(c+n)} \zeta^n,
\label{e4.20}
\end{eqnarray}
to obtain
\begin{eqnarray}
1\sim {\displaystyle\left[{a_{\rm c}-1\over a_{\rm c}}+{a-a_{\rm c}\over
a_{\rm c}^2}+{\eta\over a_{\rm c}}\right]} \left[{D-1\over D-2}
\left(1+\eta{2\over 4-D}\right)+\eta^{{D\over 2}-1} {K\over 2-D}\right],
\label{e4.21}
\end{eqnarray}
where
\begin{eqnarray}
K={2^{D\over 2}\over \sqrt{\pi}} \Gamma\left({D+1\over 2}\right)
\Gamma\left(2-{D\over 2}\right),
\label{e4.21aaa}
\end{eqnarray}
which is valid near the critical point $(a_{\rm c},z_{\rm c}=1)$. Note that
the value of $a_{\rm c}$ depends on $D$ and must be determined by
Eq.~(\ref{e4.21}).

Our results are as follows. Leading-order asymptotic analysis for small
$\eta$ gives the location of the critical point $(a_{\rm c},z_{\rm c})$; the
critical point lies at $(1,1)$ for $0<D\leq 2$ and at $(D-1,1)$ for $D\geq
2$. A next-order asymptotic analysis of Eq.~(\ref{e4.21}) for the case
$0<D<2$ yields \begin{eqnarray}
z(a)\sim 1-\left({K\over 2-D}\right)^{2\over 2-D} (a-a_{\rm c})^{2\over 2-D}
\quad (a\to a_{\rm c}^+)
\label{e4.22}
\end{eqnarray}
and
\begin{eqnarray}
{}F(a)\sim \left({K\over 2-D}\right)^{2\over 2-D}(a-a_{\rm c})^{D\over 2-D}
\quad (a\to a_{\rm c}^+).
\label{e4.23}
\end{eqnarray}
Equation (\ref{e4.23}) reduces to Eq.~(\ref{e3.10}) when $D=1$.

{}For the case $2<D<4$ a next-order asymptotic analysis of Eq.~(\ref{e4.21})
gives
\begin{eqnarray}
z(a)\sim 1-\left[(D-2) K\right ]^{2\over D-2}(a-a_{\rm c})^{2\over D-2} 
\quad (a\to a_{\rm c}^+)
\label{e4.24}
\end{eqnarray}
and
\begin{eqnarray}
{}F(a)\sim \left [(D-2)K \right ]^{2\over D-2} (a-a_{\rm c})^{2\over D-2}
\quad (a\to a_{\rm c}^+).
\label{e4.25}
\end{eqnarray}

When $D>4$ we find that
\begin{eqnarray}
z(a)\sim 1 - {D-4\over D(D-1)} (a-a_{\rm c})\quad (a\to a_{\rm c}^+)
\label{e4.26}
\end{eqnarray}
and
\begin{eqnarray}
{}F(a)\sim {D-4\over D(D-1)(D-2)} (a-a_{\rm c})\quad (a\to a_{\rm c}^+).
\label{e4.27}
\end{eqnarray}

The special case $D=3$ can be solved exactly in closed form:
\begin{eqnarray}
z(a)=1 - {(a-2)^2\over a^2+4}\quad (a\geq a_{\rm c}=2)
\label{e4.26D=3}
\end{eqnarray}
and
\begin{eqnarray}
{}F(a)= {(a-2)^2\over a^3} \quad (a\geq a_{rm c}=2).
\label{e4.27D=3}
\end{eqnarray}
Indeed, the difference equation (\ref{e4.1}) can be solved exactly and in closed
form for {\sl all} odd-integer $D$; the solution that vanishes as $n\to\infty$
is given by
$$g_n=r_-^n {\cal P}_{D-1\over 2}(n),$$
where ${\cal P}_k(n)$ is a polynomial in the variable $n$ of degree $k$.
When $D$ is an odd integer the hypergeometric series in Eq.~(\ref{e4.18})
truncates for $D\geq 5$. Unfortunately, except for the cases $D=1$ and $D=3$
we do not obtain a {\sl simple} form for the solution for $z(a)$ and $F(a)$.
An implicit solution for $z(a)$ when $D=5$, for example, is given by
$$(9a^2+64)z^3-56az^2-(8a^2+48)z+48a=0.$$

The special cases $D=0$, $D=2$, and $D=4$ need to be treated separately. For
$D=0$ the eigenvalue condition Eq.~(\ref{e4.18}) becomes very simple because
we can use the identity
\begin{eqnarray}
{}_2F_1(a,b;a;\zeta)=(1-\zeta)^{-b}.
\nonumber
\end{eqnarray}
Elementary algebra then yields
\begin{eqnarray}
z(a)={1\over a}
\label{e4.27y}
\end{eqnarray}
for all $a$. Thus, the boundary $B_2$ is a hyperbola for all $a$; the
straight-line portion of $B_2$ for $a<1$ disappears. To understand this
result observe that when $D=0$ Eq.~(\ref{e4.1}) states that random walkers
in $region~2$ cannot move outward. The appearance of this restriction is an
artifact of the uniform approximation in Eq.~(\ref{e1.8}). Thus, a steady-state 
solution has $g_n=0$ for $n>2$ and consists of random walkers
oscillating  between $region~1$ and $region~2$. In this case, the fraction
$F(a)$ of walkers in $region~1$ is exactly
\begin{eqnarray}
{}F(a)={1\over 1+a}.
\label{e4.27z}
\end{eqnarray}
{}For this degenerate case there is no critical point and no phase
transition. We emphasize that the disappearance of a phase transition is an
artifact; the uniform approximation in Eq.~(\ref{e1.8}) is only valid when
$D>0$.

{}For $D=2$ we find that
\begin{eqnarray}
z(a)\sim 1-({\rm constant})~e^{-{2\over a-a_{\rm c}}} \quad (a\to a_{\rm
c}^+) \label{e4.28}
\end{eqnarray}
and
\begin{eqnarray}
{}F(a)\sim {{\rm constant}\over a-a_{\rm c}} e^{-{2\over a-a_{\rm c}}} \quad
(a\to a_{\rm c}^+).
\label{e4.29}
\end{eqnarray}
{}For $D=4$ we have
\begin{eqnarray}
z(a)\sim 1- {a-a_{\rm c}\over 6\log\left ({1\over a-a_{\rm c}} \right
)}\quad (a\to a_{\rm c}^+)
\label{e4.30}
\end{eqnarray}
and
\begin{eqnarray}
{}F(a)\sim {a-a_{\rm c}\over 12\log \left ({1\over a-a_{\rm c}}\right)} \quad
(a\to a_{\rm c}^+).
\label{e4.31}
\end{eqnarray}
The results in Eqs.~(\ref{e4.22}-\ref{e4.31}) confirm the formulas given in 
Eqs.~(\ref{e1.11}-\ref{e1.13}).

The limiting case $D\to\infty$ is interesting because, like the case $D=1$,
we can find the exact equation for the curved portion of $B_2$. To treat
this case we perform a large-$D$ asymptotic expansion of the integral in the 
the eigenvalue condition given in Eq.~(\ref{e4.17}). Using Laplace's method
we obtain an asymptotic expansion of this condition as a formal series in
powers of $1/D$. We recover from this condition an expression for $z$ as a
function of $a/a_{\rm c}$:
\begin{eqnarray}
z(a)\sim {a_{\rm c}\over a}+{2\over D}\left [{a_{\rm c}\over a}-\left ({a_
{\rm c} \over a}\right)^3\right ]+{\cal O}\left(D^{-2}\right)\quad
(D\to\infty). \label{e4.325}
\end{eqnarray}
As one can see from Eq.~(\ref{e4.27}), in the limit $D\to\infty$ the
transition at $a=a_{\rm c} = D-1$ is still second order. However, in this
limit the discontinuity in the slope of $F(a)$ disappears and $F(a)\to 0$
for all $a$.

Equations (\ref{e4.24}) and (\ref{e4.26}) indicate that there is a change in the
form of the transition at $D=4$. When $D<4$ the slope of the boundary curve
$B_2$ is continuous, and the critical exponent depends on $D$. However, when
$D>4$ an elbow appears in $B_2$ at the critical value $a_{\rm c}=D-1$, and the
critical exponent is independent of $D$. Specifically, when $D>4$ the slope of
$B_2$ is $0$ for $0\leq a<D-1$; just above $a=D-1$ the slope abruptly becomes
$-{D-4\over D(D-1)}$.

We conclude this section by presenting a quick heuristic argument that
reproduces the results in Eqs.~(\ref{e4.26}-\ref{e4.27}). For the case $D>4$
we showed in Ref.~\cite{BeCoMe} that $T_1(D)$, the expected time for a
random walker who originates in $region~1$ to return to $region~1$, is given
by \begin{eqnarray}
T_1(D)=2{D-2\over D-4}.
\label{e4.32}
\end{eqnarray}
In a steady state all $g_1$ random walkers in $region~1$ leave this region
and in a $z=1$ model only the fraction $\Pi_1(D)$ ever return. The random
walkers who return to $region~1$ do so in $T_1(D)$ steps on average. These
returning random walkers experience a death rate $z$ for $T_1(D)-1$ of these
$T_1(D)$ steps. Thus, the expected number of random walkers who actually
return to $region~1$ is decreased by the factor $z^{T_1(D)-1}$. Hence, after
$T_1(D)$ steps we expect to find $a\Pi_1(D) z^{T_1(D)-1}g_1$ random walkers
in $region~1$. The condition that there be a steady state is therefore given
by \begin{eqnarray}
a\Pi_1(D) z^{T_1(D)-1} =1.
\label{e4.33}
\end{eqnarray}
Using the expressions for $\Pi_1(D)$ and $T_1(D)$ in Eqs.~(\ref{e2.6b}) and
(\ref{e4.32}) we obtain an approximate relation between $z$ and $a$ that is
valid near the critical point; that is, where $a=D-1+\delta$, $z=1-\epsilon$
as $\delta,\epsilon\to 0^+$. To first order in $\delta$ and $\epsilon$ this
approximate relation is
\begin{eqnarray}
\epsilon \sim \delta {D-4\over D(D-1)},
\label{e4.34}
\end{eqnarray}
which is precisely the result in Eq.~(\ref{e4.26}). We obtain the result in
Eq.~(\ref{e4.27}) by substituting Eq.~(\ref{e4.26}) into Eq.~(\ref{e4.3}).
Note that this argument is valid only for $a\geq a_{\rm c}=D-1$.

While the above argument is only valid in the neighborhood of $a_{\rm c}$,
we can also use the above reasoning to derive the entire curve $z(a)$ in the
limit $D\to\infty$. In this limit, $T_1=2$. Hence, from Eq.~(\ref{e4.33}) we
have \begin{eqnarray}
z={a_{\rm c}\over a},
\label{e4.35}
\end{eqnarray}
the leading behavior in Eq.~(\ref{e4.325}).

\section*{ACKNOWLEDGEMENTS}
\label{s5}
Two of us, CMB and PNM, wish to thank the U.S. Department of Energy for
financial support under grant number DE-FG02-91-ER40628. SB also thanks the
U.S. Department of Energy for support under grant number DE-AC02-76-CH00016.

\appendix{LARGE-TIME ASYMPTOTIC BEHAVIOR \label{a1}}

In this Appendix we analyze the large-$t$ behavior of the distribution
$G_{n,t}$ of random walkers for the uniform approximation of the
probabilities $P_{\rm out}$ and $P_{\rm in}$ given in Eq.~(\ref{e1.8}). To
this end we solve the set of equations in (\ref{e1.10}) for the Kronecker
delta initial condition $G_{n,0}=\delta_{n,1}$. As discussed earlier, the
large-$t$ behavior of a dissipative process is independent of the specific
choice of initial condition.

{}First, we derive a formal solution for $G_{n,t}$ that is valid for general
$P_{\rm out}$ and $P_{\rm in}$. We define 
\begin{equation}
d_{n,t}=\cases{
a z^{n-2} \left[\prod_{i=1}^{n-1} P_{\rm out}(i)\right] G_{n,t} \quad & $(n\geq 2)$,\cr
\noalign{\medskip}
G_{1,t} & $(n=1)$,\cr}
\label{eA.1}
\end{equation}
and rewrite Eqs.~(\ref{e1.10}) as
\begin{equation}
d_{n,t}=\cases{Q_n d_{n+1,t-1}+d_{n-1,t-1} \quad & $(n\geq 2)$,\cr 
\noalign{\medskip}
Q_1 d_{2,t-1}\quad & $(n=1)$,\cr}
\label{eA.2}
\end{equation}
where we let
\begin{equation}
Q_n=\cases{z^2 P_{\rm out}(n) P_{\rm in}(n+1) \quad & $(n\geq 2)$,\cr 
\noalign{\medskip}
a z P_{\rm out}(1) P_{\rm in}(2) \quad & $(n=1)$.\cr}
\label{eA.3}
\end{equation}
Next, we define the generating function 
\begin{equation}
e_n(y)=\sum_{t=0}^{\infty} d_{n,t} y^t
\label{eA.4}
\end{equation}
and obtain from Eqs.~(\ref{eA.2})
\begin{equation}
e_n(y)=\cases{y Q_n e_{n+1}(y)+y e_{n-1}(y) \quad & $(n\geq 2)$,\cr
\noalign{\medskip}
1+y Q_1 e_2(y) \quad & $(n=1)$,\cr}
\label{eA.5}
\end{equation}
where we have applied the Kronecker delta initial condition.

Let us define a continued fraction by the recursion relation
\begin{equation}
S_n(y^2)={1\over 1-y^2 Q_n S_{n+1}(y^2)} \quad (n\geq 1).
\label{eA.6}
\end{equation}
It is easy to show that for $n\geq 3$ the recursion relation in
Eq.~(\ref{eA.5}) is satisfied by
\begin{equation}
e_n(y)=A~y^{n-1}~\prod_{i=2}^n S_i(y^2) \quad (n\geq 2).
\label{eA.7}
\end{equation}
[Since $e_n(y)$ obeys a second-order difference equation, there is a linearly
independent solution which can be determined using the technique of variation of
parameters. This solution does not contribute; apparently, it fails to obey the
appropriate boundary conditions at $n=\infty$.] We determine $e_1$ and the
constant $A$ by solving simultaneously the special cases $n=1$ and $n=2$ of
Eqs.~(\ref{eA.5}):
\begin{eqnarray}
A y S_2(y^2)&=&A Q_2 y^3 S_2(y^2) S_3(y^2) + y e_1(y),\nonumber\\
e_1(y)&=&1+A y^2 Q_1 S_2(y^2).
\label{eA.8}
\end{eqnarray}
Solving the above equations leads to a surprisingly compact expression for
all $e_n(y)$:
\begin{equation}
e_n(y)=y^{n-1}~\prod_{i=1}^n S_i(y^2) \quad (n\geq 1).
\label{eA.9}
\end{equation}
Using a contour integral to project out the coefficients in the generating
function we obtain
\begin{equation}
G_{n,t}=\left({a\over z}\right)^{1-\delta_{1,n}} \left[\prod_{i=1}^{n-1} z
P_{\rm out}(i)\right]
\oint_C {dy\over 2\pi i y} y^{n-t-1} \prod_{i=1}^{n} S_i(y^2), \label{eA.10}
\end{equation}
where empty products are defined to be unity. The contour $C$ encircles the
pole at the origin in the complex-$y$ plane but excludes all other
singularities of the integrand.

This rather strange expression (a contour integral over a product of
continued fractions!) is of little use, even in an asymptotic analysis for
large values of $t$. Only for particular choices for $P_{\rm out}$ and
$P_{\rm in}$ is progress possible. A significant advantage of the uniform
approximation in Eq.~(\ref{e1.8}) [compared with probabilities given in
Eqs.~(\ref{e1.5}-\ref{e1.7})] is that they simplify the expression for
$G_{n,t}$ in Eq.~(\ref{eA.10}) for {\sl all} $D>0$. [The probabilities in
Eqs.~(\ref{e1.5}-\ref{e1.7})] lead to a tractable result only for $D=0,~1,$
and $2$.]

We simplify the expression for $G_{n,t}$ in Eq.~(\ref{eA.10}) by recalling
the continued fraction representation for a hypergeometric function
\cite{BO} : \begin{eqnarray}
{{}_2F_1(a,b+1;c+1;\zeta)\over {}_2F_1(a,b;c;\zeta)}=1/(1+f_1 \zeta/(1+f_2 \zeta
/(1+f_3 \zeta/(1+\ldots)))),
\label{eA.11}
\end{eqnarray}
where
\begin{eqnarray}
f_{2i}=-{\displaystyle{(i+b)(i+c-a)\over(2i+c)(2i+c-1)}}\quad {\rm and}\quad
f_{2i+1}=-{\displaystyle{(i+a)(i+c-a)\over (2i+c)(2i+c+1)}}.
\label{eA.11aaaaaaaaaa}
\end{eqnarray}
Substituting the uniform approximation for $P_{\rm out}$ and $P_{\rm in}$ in
Eq.~(\ref{e1.8}) into Eq.~(\ref{eA.3}) gives
\begin{equation}
Q_n=z^2{n(n+D-2)\over (2n+D-3)(2n+D-1)}\qquad (n\geq 2),
\label{eA.12}
\end{equation}
which can be rewritten as
\begin{eqnarray}
Q_{n+2i}&=z^2 {\displaystyle{(i+{n+D-2\over 2})(i+{n\over 2})\over (2i+n+{D-
3\over 2})(2i+n+{D-1\over 2})}}\quad (n\geq 2, i\geq 1),\nonumber\\
\noalign{\medskip}
Q_{n+2i+1}&=z^2 {\displaystyle{(i+{n+D-1\over 2})(i+{n+1\over 2})\over
(2i+n+{D-1\over 2})(2i+n+{D+1\over 2})}}\quad (n\geq 2,i\geq 0).
\label{eA.13}
\end{eqnarray}
The continued fractions in Eq.~(\ref{eA.6}) can thus be identified as
\begin{equation}
S_n(y^2)={{}_2F_1\left({n\over 2},{n+1\over 2};n+{D-1\over 2};z^2 y^2\right)
\over {}_2F_1\left({n\over 2},{n-1\over 2};n+{D-3\over 2};z^2 y^2\right)}\quad
(n\geq 2).
\label{eA.14}
\end{equation}
Hypergeometric functions are symmetric in their first two arguments.
Therefore, \begin{equation}
\prod_{i=2}^n S_i(y^2)={{}_2F_1\left({n\over 2},{n+1\over 2};n+{D-1\over 2};z^2
y^2 \right)\over {}_2F_1\left({1\over 2},1;{D+1\over 2};z^2 y^2\right)}\quad
(n\geq 2). \label{eA.15}
\end{equation}
Substituting this last result into Eq.~(\ref{eA.10}) we finally obtain
\begin{equation}
G_{n,t}=z^t~{\Gamma\left({D+1\over 2}\right)\Gamma(n+D-2)\over 2^{n-2}
\Gamma(D) \Gamma\left(n+{D-3\over 2}\right)} \oint_C {dy\over 2\pi i
y}~y^{n-t-1}{\left( {z\over a}\right)^{\delta_{1,n}} {}_2F_1\left({n\over
2},{n+1\over 2};n+{D-1\over 2}; y^2\right)\over 1+\left({z\over a}-1\right)
{}_2F_1\left({1\over 2},1;{D+1\over 2}; y^2\right)}.
\label{eA.16}
\end{equation}

Recalling Eq.~(\ref{e2.1}) we obtain an expression for the total number of
walkers at time $t$ by summing Eqs.~(\ref{e1.10}) over all positive integers
$n$:
\begin{equation}
N_t=z^t~\left[1+\left({a\over z}-1\right) \sum_{\tau=0}^{t-1} z^{-\tau}
G_{1,\tau}\right].
\label{eA.17}
\end{equation}
Next we insert $G_{1,t}$ from Eq.~(\ref{eA.16}) and sum over $\tau$:
\begin{equation}
N_t=z^t~\left[1+\left(1-{z\over a}\right) \oint_C {dy~y^{-t}\over 2\pi i (1-
y)} {{}_2F_1\left({1\over 2},1;{D+1\over 2};y^2\right)\over 1+\left({z\over
a}-1\right) {}_2F_1\left({1\over 2},1;{D+1\over 2};y^2\right)}\right],
\label{eA.18}
\end{equation}
where we have eliminated terms in the integrand that are regular at the origin
in the complex-$y$ plane. From $G_{1,t}$ in Eq.~(\ref{eA.16}) and $N_t$ in
Eq.~(\ref{eA.18}) we obtain the large-$t$ behavior of the fraction $F_t$ in
Eq.~(\ref{e2.2}). Note the similarity of the denominator in both integrals with
the eigenvalue condition in (\ref{e4.185}) for steady-state solutions. The
asymptotic behavior of the integrals for large $t$ is dominated by the poles of
the integrands, and the steady-state solution is merely the special case where
the asymptotic behavior is independent of $t$ in leading order.

To extract the large-$t$ behavior of $N_t$ and $F_t$, we conduct a saddle-point 
analysis of the contour integrals for $G_{1,t}$ and $N_t$ in Eqs.~(\ref{eA.16})
and (\ref{eA.18}). Both expressions, aside from the prefactor $z^t$, only depend
on the ratio $a/z$. Saddle-point analysis requires that we consider three
distinct cases, values of $a/z$ such that (i) $(a,z)$ lies to the left of the
line $B_1$, (ii) $(a,z)$ is on $B_1$, and (iii) $(a,z)$ lies to the right of
$B_1$ (see Fig.~\ref{f1}). For all cases the integrands for both $G_{1,t}$ and
$N_t$ have a pole at $y=y_{\rm p}$ on the real positive axis. For case (i)
$y_{\rm p}>1$, for case (ii) $y_{\rm p}=1$, and for case (iii) $y_{\rm p}<1$. 

{}For case (i) we find to leading order that as $t\to\infty$
\begin{equation}
G_{1,t}\sim\cases{{\displaystyle
{{a\over z}\over\left(1-{a\over z}\right)^2} {(2-D)^2\over 2 K \Gamma
\left({D\over 2}\right)} {z^t t^{{D\over 2}-2}} } \quad & $(0<D<2)$,\cr
\noalign{\medskip}
{\displaystyle {{a\over (D-1) z}\over\left(1-{a\over (D-1)z}\right)^2}
{2^{{D \over 2}-2} (D-2)^2 \Gamma\left({D-1\over 2}\right)\over \sqrt{\pi}}
{z^t t^{-{D\over 2}}} } \quad & $(D>2)$,\cr}
\label{eA.19}
\end{equation}
and
\begin{equation}
N_t\sim\cases{{\displaystyle {{a\over z}\over\left(1- {a\over z}\right)}
{(2-D) \over K \Gamma\left({D\over 2}\right)} {z^t t^{{D\over 2}-1}} } \quad
& $(0<D<2)$,\cr
\noalign{\medskip}
{\displaystyle {{a\over (D-1) z}\over 1-{a\over (D-1) z}} (D-2) z^t } \quad
& $(D>2)$,\cr}
\label{eA.20}
\end{equation}
where $K$ is given in Eq.~(\ref{e4.21aaa}). Hence,
\begin{equation}
{}F_t\sim\cases{{\displaystyle{1-{D\over 2}\over 1-{a\over z}} t^{-1}}\quad &
$(0<D<2)$,\cr
\noalign{\medskip}
{\displaystyle{1\over 1-{a\over (D-1) z}}{2^{{D\over
2}-2}(D-2)\Gamma\left({D-1 \over 2}\right)\over \sqrt{\pi}}t^{-{D\over 2}}
} \quad & $(D>2)$.\cr}  \label{eA.21}
\end{equation}

{}For case (ii) we find to leading order that as $t\to\infty$
\begin{equation}
G_{1,t}\sim\cases{{\displaystyle {K\over 2 \Gamma\left(2-{D\over 2}\right)}
z^t t^{-{D\over 2}} } \quad & $(0<D<2)$,\cr
\noalign{\medskip}
{\displaystyle {D-1\over 2 K \Gamma\left({D\over 2}\right)} z^t t^{{D\over
2}-2} } \quad & $(2<D<4)$,\cr
\noalign{\medskip}
{\displaystyle {D-4\over 2 (D-2)} z^t} \quad & $(D>4)$,\cr}
\label{eA.22}
\end{equation}
and
\begin{equation}
N_t\sim\cases{{\displaystyle z^t} \quad & $(0<D<2)$,\cr
\noalign{\medskip}
{\displaystyle {D-1\over K \Gamma\left({D\over 2}\right)} z^t~t^{{D\over
2}-1}} \quad & $(2<D<4)$,\cr
\noalign{\medskip}
{\displaystyle {D-4\over 2} z^t t} \quad & $(D>4)$.\cr}
\label{eA.23}
\end{equation}
Hence,
\begin{equation}
{}F_t\sim\cases{{\displaystyle {D-1\over K\Gamma\left({D\over 2}\right)} t^{-
{D\over 2}} } \quad & $(0<D<2)$,\cr
\noalign{\medskip}
{\displaystyle {1\over 2t}} \quad & $(2<D<4)$,\cr
\noalign{\medskip}
{\displaystyle {1\over (D-2) t}} \quad & $(D>4)$.\cr} 
\label{eA.24}
\end{equation}

The analysis of case (iii) is somewhat more complicated. The saddle point in
cases (i) and (ii) is very near $y=1$ for large $t$, but in case (iii) the
integrands have poles at $0<y=y_{\rm p}(a/z)<1$ and the saddle point is now
located near $y_{\rm p}$. An asymptotic analysis of this case for large $t$ is
possible only if we consider a small neighborhood to the right of the line
$B_1$. Approaching $B_1$ we find that $y_{\rm p}\to 1^-$. We use the ansatz
${a\over z}={a_{\rm c}\over z_{\rm c}}+\epsilon$ and $y_{\rm p}(a/z)=1-\delta(
\epsilon)$, where $\epsilon\ll 1$ and $\delta\ll 1$, but where $t\delta\gg 1$.
We find that as $t\to\infty$
\begin{equation}
\delta(\epsilon)\sim\cases{{\displaystyle \epsilon^{2\over 2-D} \left[K\over
2-D\right]^{2\over 2-D}}\quad & $(0<D<2)$,\cr
\noalign{\medskip}
{\displaystyle \epsilon^{2\over D-2}\left[K (D-2)\right]^{-{2\over D-2}}
}\quad & $(2<D<4)$,\cr 
\noalign{\medskip}
{\displaystyle \epsilon {D-4\over 2 (D-1)(D-2)}} \quad & $(D>4)$,\cr}
\label{eA.255}
\end{equation}
leading to
\begin{equation}
G_{1,t}\sim\cases{{\displaystyle \epsilon^{D\over 2-D}~{2 \over 2-
D}\left[K\over 2-d\right]^{2\over 2-D} \left[{z\over y_{\rm
p}(a/z)}\right]^t }\quad & $(0<D<2)$,\cr
\noalign{\medskip}
{\displaystyle \epsilon^{4-D\over D-2}~{2(D-1)\over D-2}\left[K
(D-2)\right]^{-{2\over D-2}} \left[{z\over y_{\rm p}(a/z)}\right]^t }\quad
& $(2<D<4)$,\cr \noalign{\medskip}
{\displaystyle {D-4\over 2 (D-2)} \left[{z\over y_{\rm p}(a/z)}\right]^t }
\quad & $(D>4)$\cr}
\label{eA.25}
\end{equation}
and
\begin{equation}
N_t\sim\cases{{\displaystyle {2\over 2-D}\left[{z\over y_{\rm
p}(a/z)}\right]^t } \quad & $(0<D<2)$,\cr
\noalign{\medskip}
{\displaystyle \epsilon^{-1}~2 (D-1) \left[{z\over y_{\rm p}(a/z)}\right]^t
} \quad & $(2<D<4)$,\cr
\noalign{\medskip}
{\displaystyle \epsilon^{-1}~( D-1)(D-2) \left[{z\over y_{\rm
p}(a/z)}\right]^t } \quad & $(D>4)$.\cr}
\label{eA.26}
\end{equation}
Hence,
\begin{equation}
{}F_t\sim\cases{{\displaystyle\epsilon^{D\over 2-D}~\left[{K\over 2-
D}\right]^{2 \over 2-D}} \quad & $(0<D<2)$,\cr 
\noalign{\medskip}
{\displaystyle \epsilon^{2\over D-2}~{\left[K(D-2)\right]^{-{2\over
D-2}}\over D-2}} \quad & $(2<D<4)$,\cr
\noalign{\medskip}
{\displaystyle \epsilon~{D-4\over 2 (D-1)(D-2)^2}} \quad & $(D>4)$.\cr}
\label{eA.27}
\end{equation}

{}From the previous formula we can recover the asymptotic results in
Eqs.~(\ref{e4.23}), (\ref{e4.25}), and~(\ref{e4.27}), which are valid on the
line $B_2$ as $a\to a_{\rm c}^+$ and $z\to z_{\rm c}^-$. The particular path
$B_2$ is distinguished merely by the fact that the total number of walkers $N_t$
approaches a nonzero constant as $t\to\infty$. Thus, the portion of $B_2$ to the
right of $B_1$ is obtained for $z=y_{\rm p}(a/z)$ in Eqs.~(\ref{eA.26}). Again,
we let $z=1-\eta$ for $\eta\to 0^+$ and find that $\eta\sim\delta(\epsilon)$.
Then, using $z_{\rm c}=1$, we find that $\epsilon\sim(a-a_{\rm c})+a_{\rm c}
\eta$. From Eq.~(\ref{eA.255}) for $D<4$, we have $\epsilon\gg\eta$, and we
merely need to identify $\epsilon=a-a_{\rm c}$ in Eq.~(\ref{eA.27}) to recover
our earlier results. For $D>4$ we recall that $\epsilon={\cal O}(\eta)$ to
recover Eq.~(\ref{e4.27}).

\figure{
Generic phase diagram for the $(a,z)$ plane. (The diagram was
actually generated using data from $D={1\over 2}$ random walks.) Shown on
the diagram are the boundary curves $B_1$ and $B_2$. To the left of $B_1$
and on $B_1$ the fraction of random walkers in $region~1$, $F_t$, approaches
$0$ as $t\to\infty$; to the right of $B_1$ this fraction approaches a finite
positive number as $t\to\infty$. Above $B_2$ the total number of random
walkers, $N_t$ diverges as $t\to\infty$; below $B_2$ the total number of
walkers approaches $0$ as $t\to\infty$. On $B_2$ the distribution of random
walkers approaches a steady state as $t\to\infty$. The critical point
$(a_{\rm c}, z_{\rm c})$ lies at the intersection of $B_1$ and $B_2$.
\label{f1}
}

\figure{
Phase diagram in the $(a,z)$ plane for the case $D=1$. For this
dimension the probabilities $P_{\rm out}(n)$ and $P_{\rm in}(n)$ for the
hyperspherical surface area case given in Eqs.~(\ref{e1.5}-\ref{e1.7}) and the uniform
approximation case given in Eqs.~(\ref{e1.8}) are the same. On this
diagram the slope of $B_1$ is unity, $a_{\rm c} = 1$, and the slope of $B_2$
is continuous.
\label{f2}
}

\figure{
Phase diagram in the $(a,z)$ plane for the case $D=2$. For this
dimension the probabilities $P_{\rm out}(n)$ and $P_{\rm in}(n)$ for the
hyperspherical surface area case and the uniform
approximation case are the same. On this
diagram the slope of $B_1$ is unity, $a_{\rm c}=1$, and the slope of $B_2$
is continuous. Note that the slope of $B_2$ approaches $0$ exponentially
fast as $a\to a_{\rm c}$ from above.
\label{f3}
}

\figure{
Phase diagram in the $(a,z)$ plane for the case $D=3$ using the
probabilities $P_{\rm out}(n)$ and $P_{\rm in}(n)$ for the uniform
approximation case in Eqs.~(\ref{e1.8}).
The slope of $B_1$ is $1/2$, $a_{\rm c}=2$, and the
slope of $B_2$ is continuous.
\label{f4}
}

\figure{
Phase diagram in the $(a,z)$ plane for the case $D=3$ using the
probabilities $P_{\rm out}(n)$ and $P_{\rm in}(n)$ for the hyperspherical
surface area case given in Eqs.~(\ref{e1.5}-\ref{e1.7}).
The slope of $B_1$ is $1-1/\zeta(2)=1-6/\pi^2$, $a_{\rm c}=2.551\ldots$, and
the slope of $B_2$ is continuous.
\label{f5}
}

\figure{
Phase diagram in the $(a,z)$ plane for the case $D=4$ using the
probabilities $P_{\rm out}(n)$ and $P_{\rm in}(n)$ for the uniform
approximation case given in Eqs.~(\ref{e1.8}). The slope of $B_1$ is $1/3$ and
$a_{\rm c}=3$. The slope of $B_2$ is continuous; it vanishes logarithmically
as $a\to a_{\rm c}$ from above [see Eq.~(\ref{e4.30})].
\label{f6}
}

\figure{
Phase diagram in the $(a,z)$ plane for the case $D=4$ using the probabilities 
$P_{\rm out}(n)$ and $P_{\rm in}(n)$ for the hyperspherical surface area case 
given in Eqs.~(\ref{e1.5}-\ref{e1.7}). The slope of $B_1$ is $1-1/\zeta(3)$
and $a_{\rm c}=5.949\ldots$. Universality arguments lead us to believe that
the slope of $B_2$ is continuous and vanishes logarithmically as $a\to
a_{\rm c}$ from above as in Fig.~\ref{f6}.
\label{f7}
}

\figure{
Phase diagram in the $(a,z)$ plane for the case $D=5$ using the
probabilities $P_{\rm out}(n)$ and $P_{\rm in}(n)$ for the uniform
approximation case given in Eqs.~(\ref{e1.8}).
The slope of $B_1$ is $1/4$ and $a_{\rm c}=4$. The
slope of $B_2$ is not continuous; there is an elbow at $a=a_{\rm c}$.
\label{f8}
}

\figure{
Phase diagram in the $(a,z)$ plane for the case $D=5$ using the
probabilities $P_{\rm out}(n)$ and $P_{\rm in}(n)$ for the hyperspherical
surface area case given in Eqs.~(\ref{e1.5}-\ref{e1.7}).
The slope of $B_1$ is $1-1/\zeta(4)=1-90/\pi^4$ and $a_{\rm
c}=13.147\ldots$. The slope of $B_2$ is not continuous; there is an elbow at
$a=a_{\rm c}$.
\label{f9}
}


\begin{references}
\bibitem{Num}
One must be careful when performing numerical studies of Eq.~(\ref{e1.10}).
This equation describes random walkers who {\sl must} move outward or inward
at every step. Thus, on an $(n,t)$ lattice the system exhibits a
checkerboard parity behavior; it is actually two noninteracting systems, one
with sites for which $n+t$ is even and the other with sites where $n+t$ is
odd. Hence, numerical studies of $F_t$ and $N_t$ with arbitrary initial
conditions in general yield oscillatory time evolution. To avoid ambiguities
one must focus on just one system. To do so one studies $N_t$ at a given
time only for even values of $n+t$ or only for odd values of $n+t$. One must
be equally careful about numerical studies of $F_t$. (In principle, one can
couple the two distinct systems by allowing walkers to have a small
probability $\epsilon$ to {\sl stay} at their current site at each time
step. This coupling would enable the systems to thermalize. Once this has
happened, one can set $\epsilon=0$ and ignore the distinctions arising from
the checkerboard parity.)
\end{references}
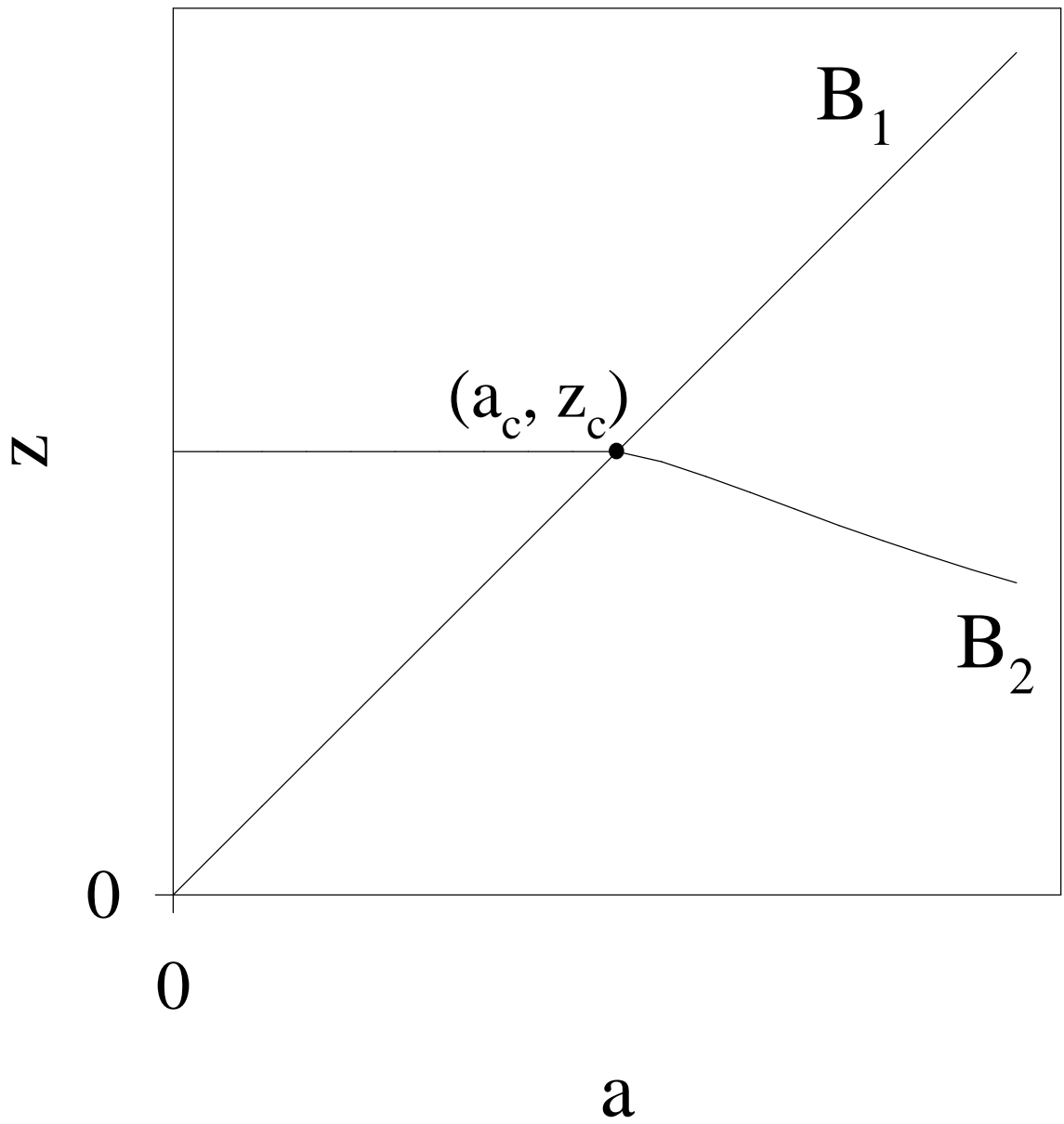
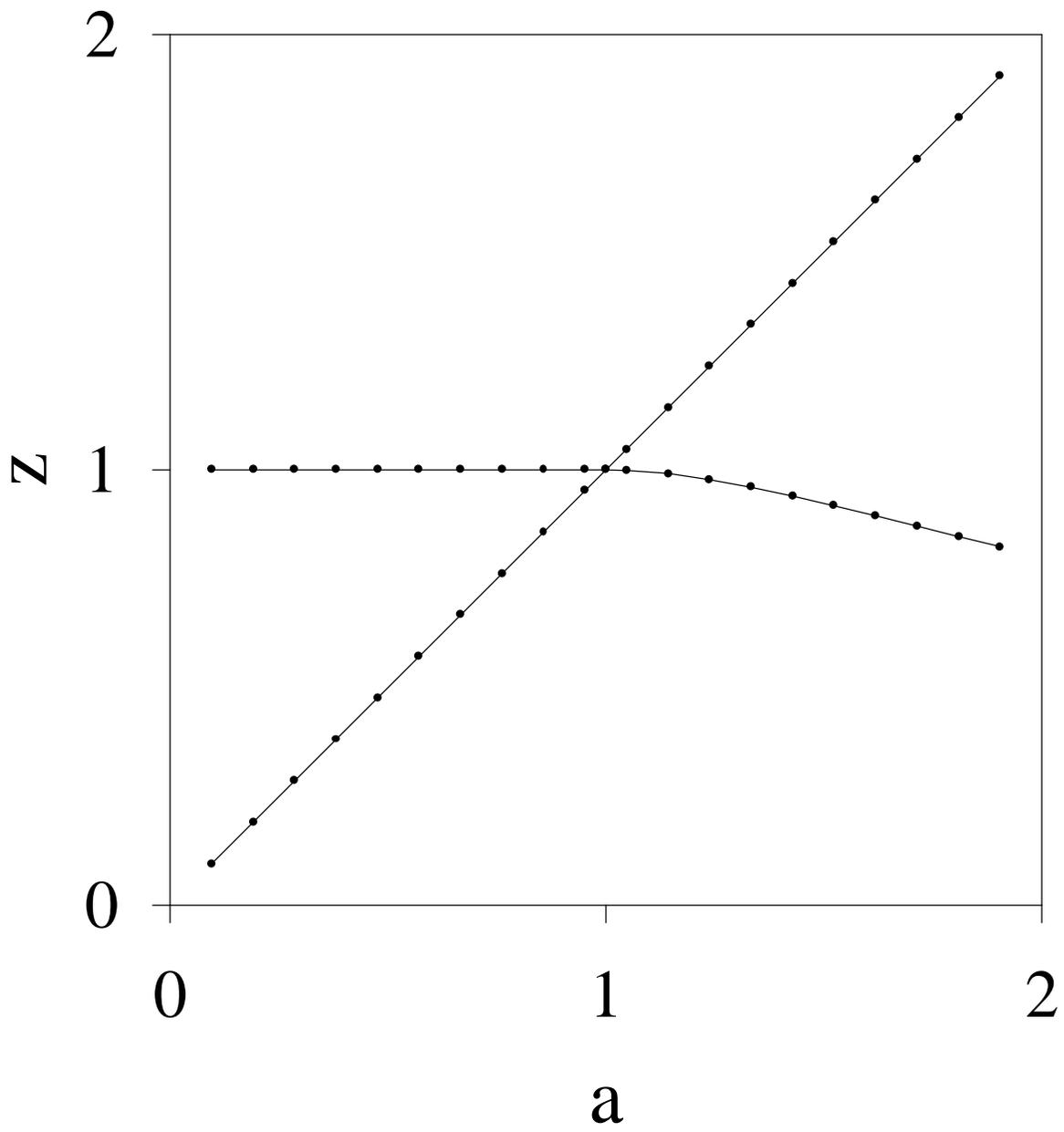
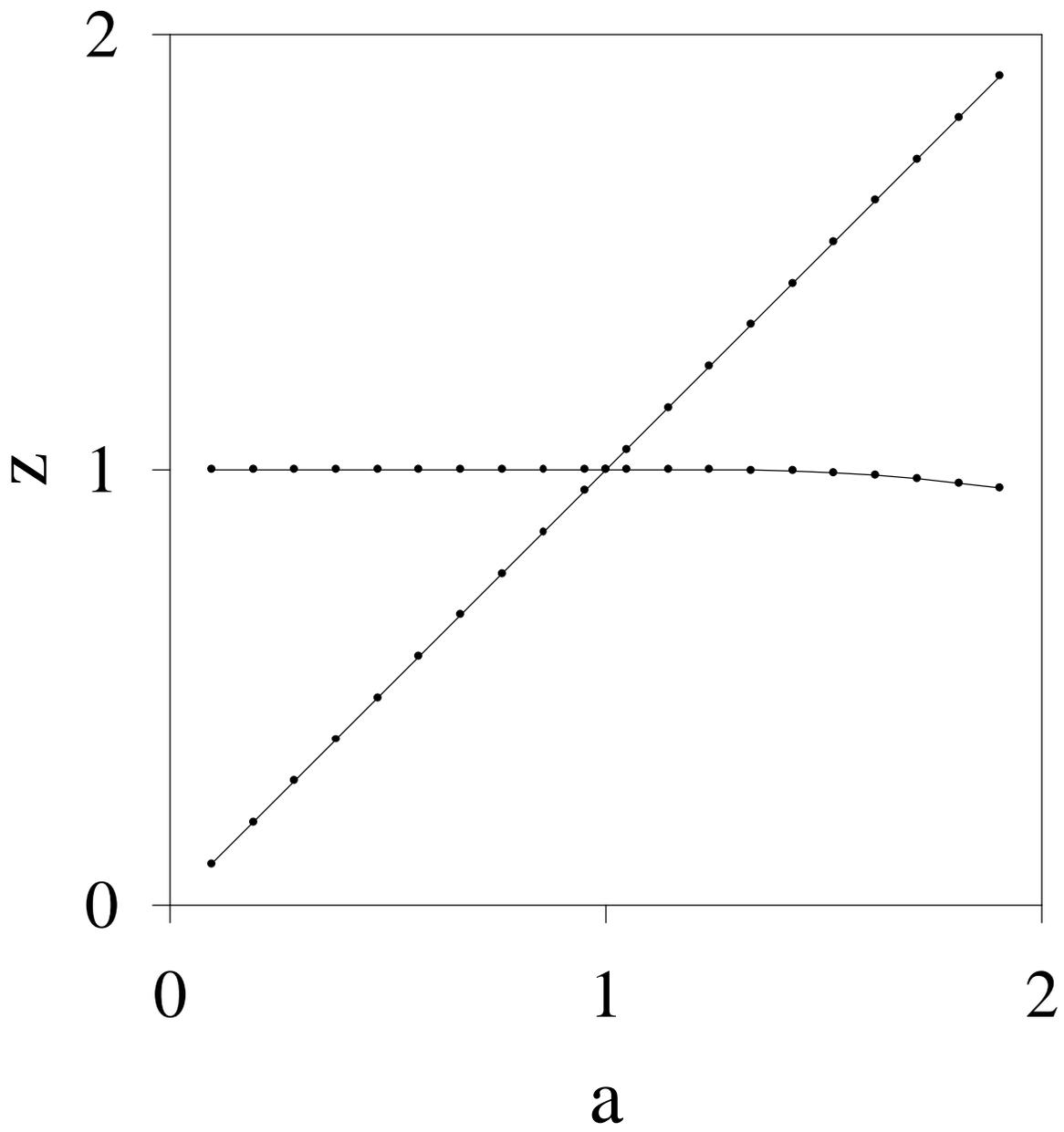
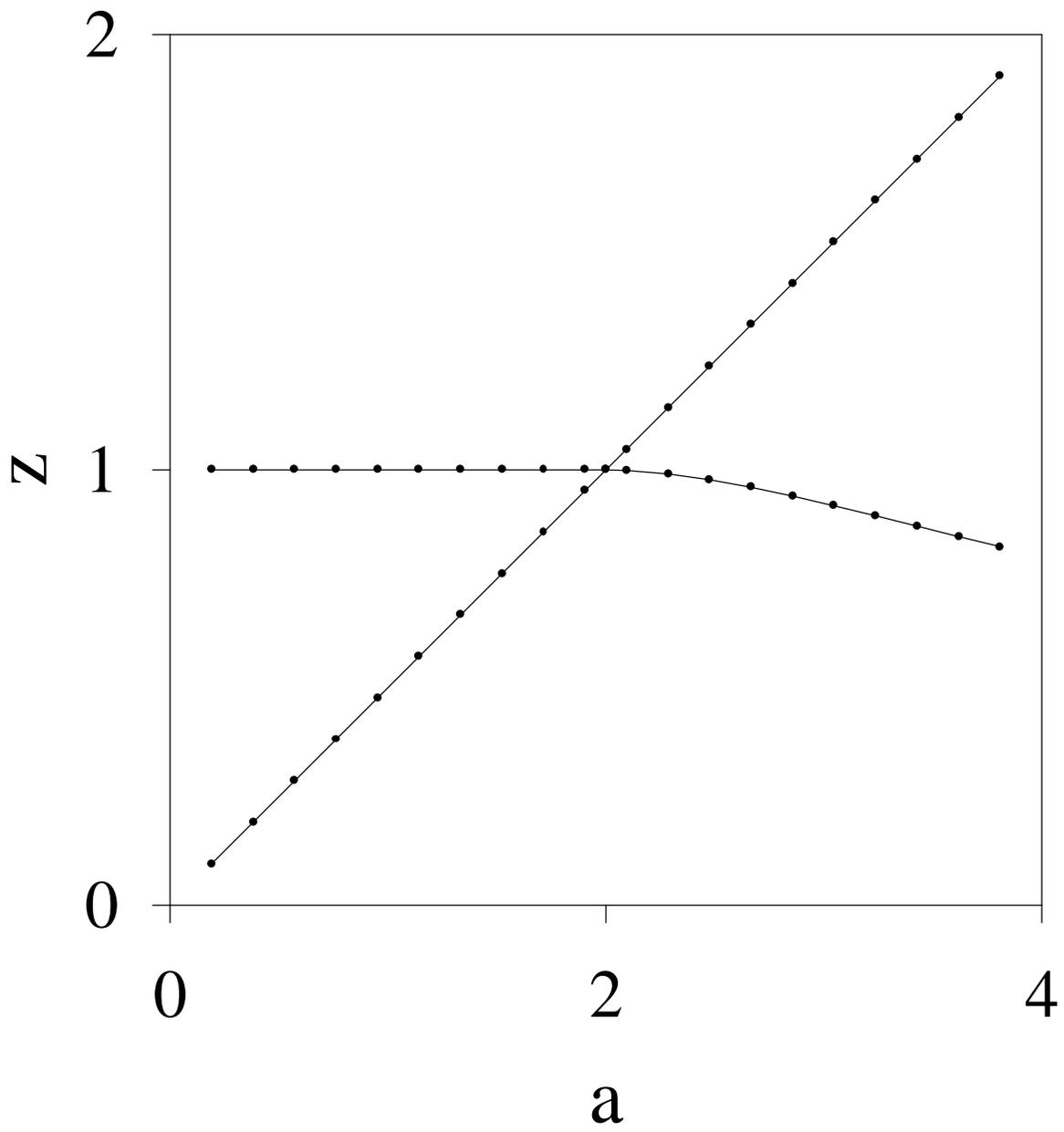
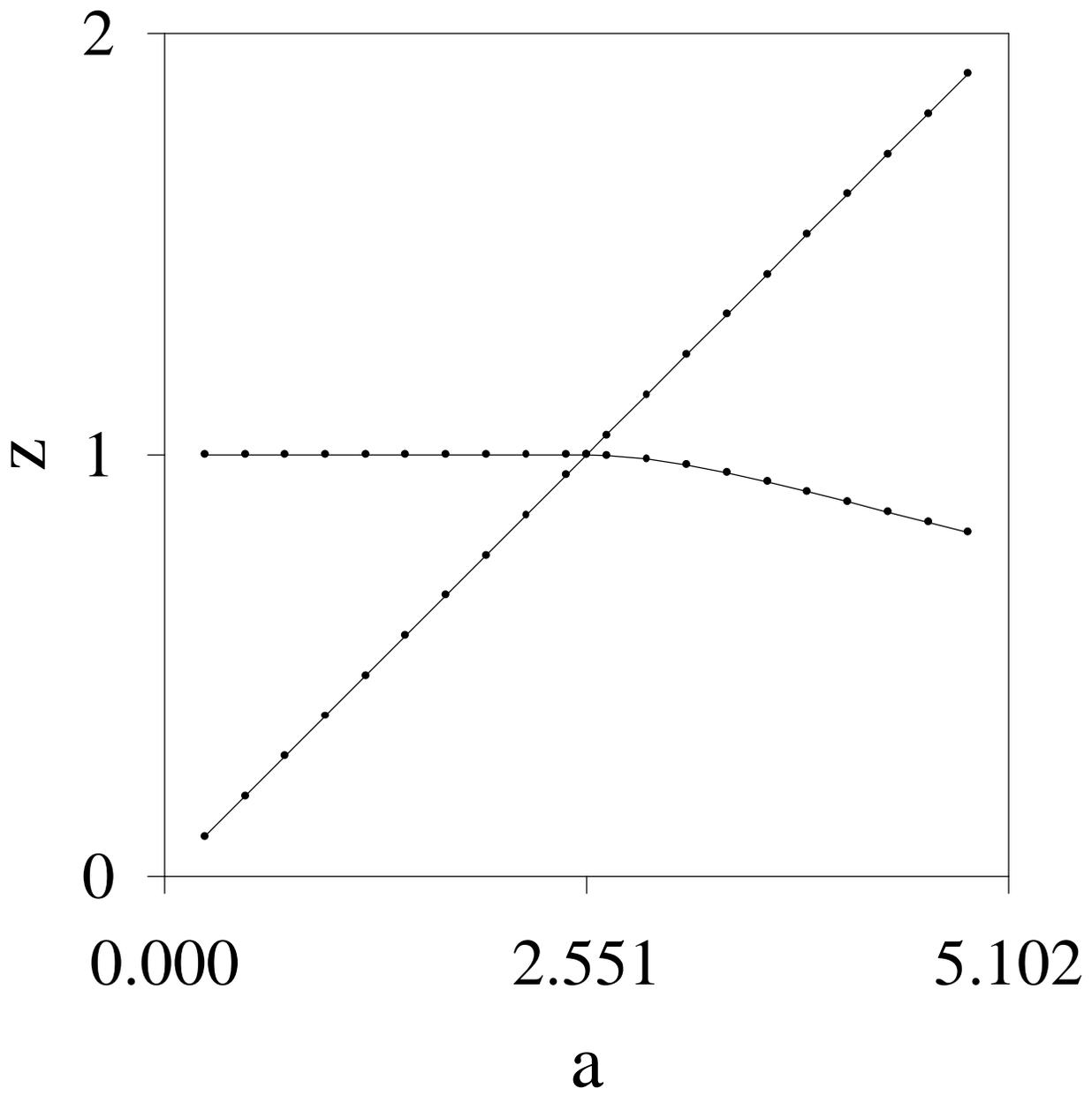
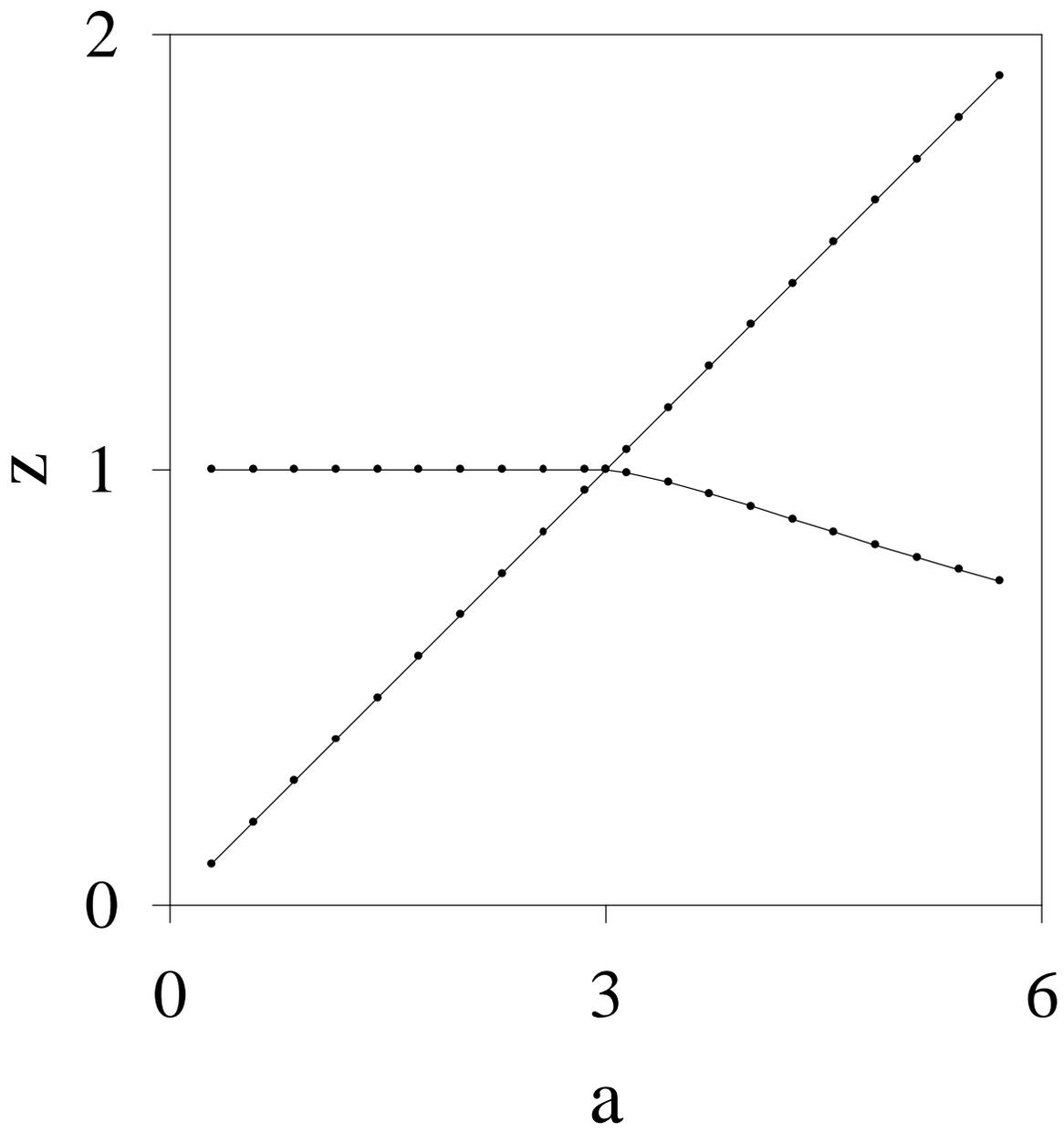
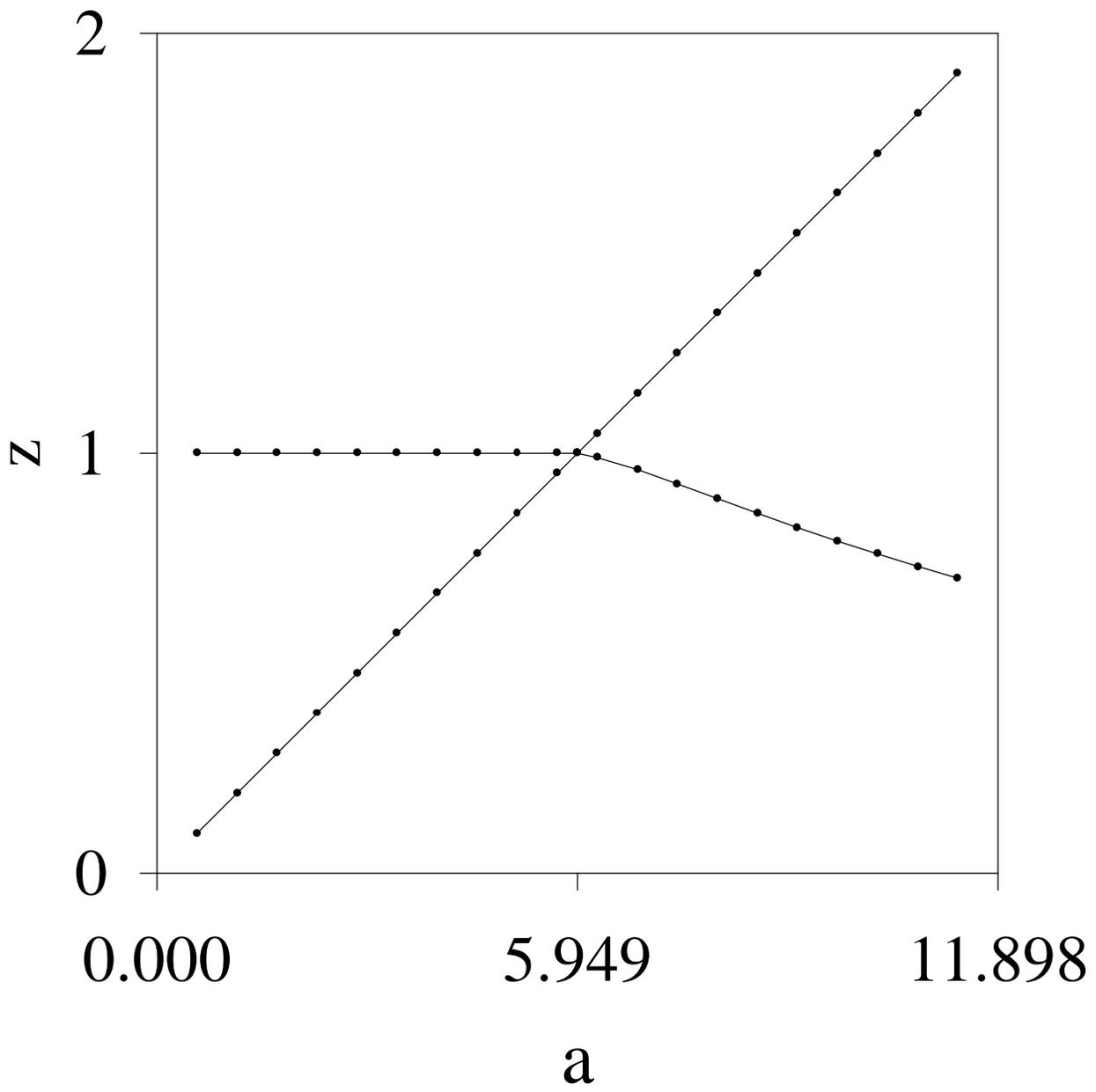
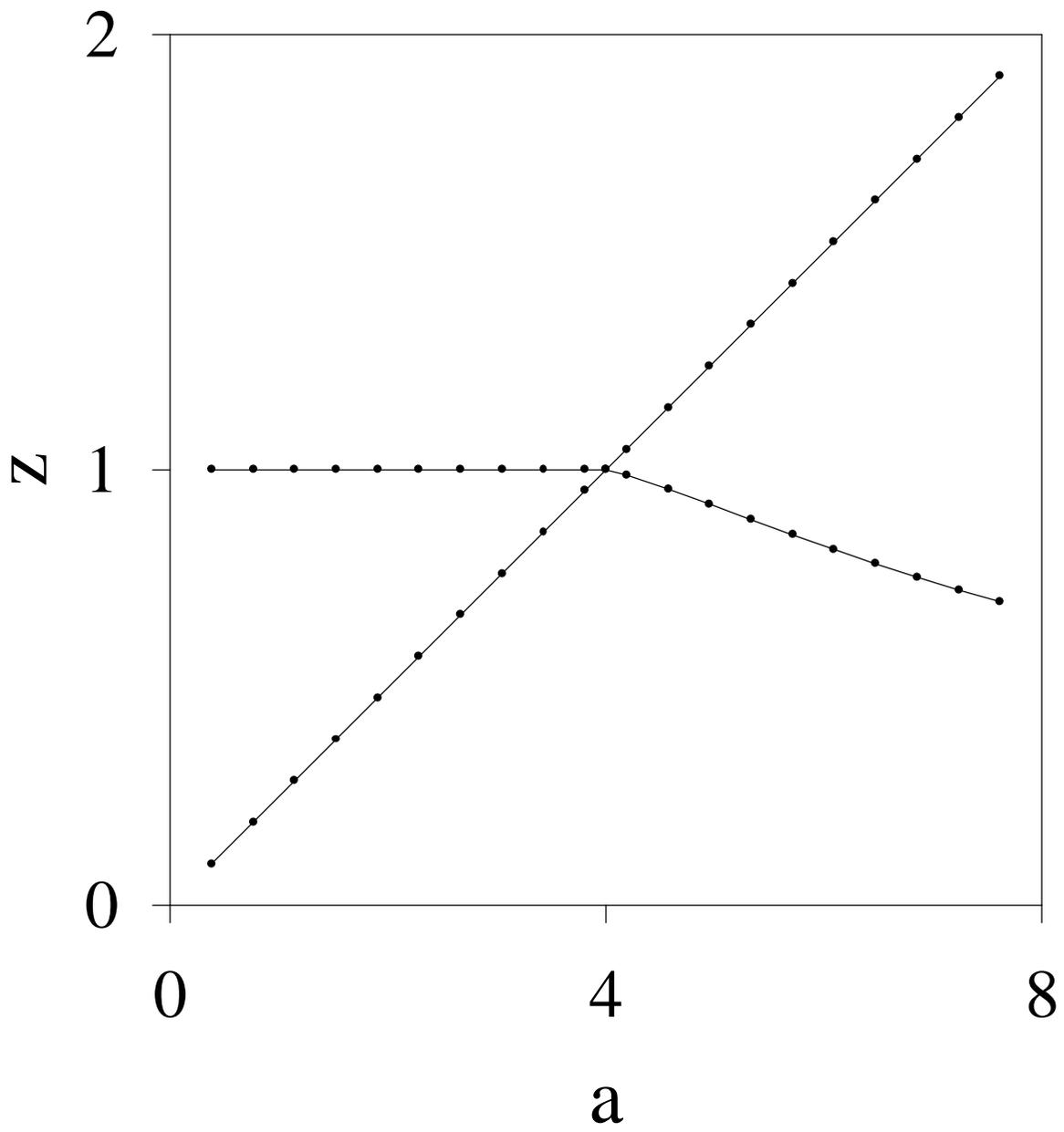
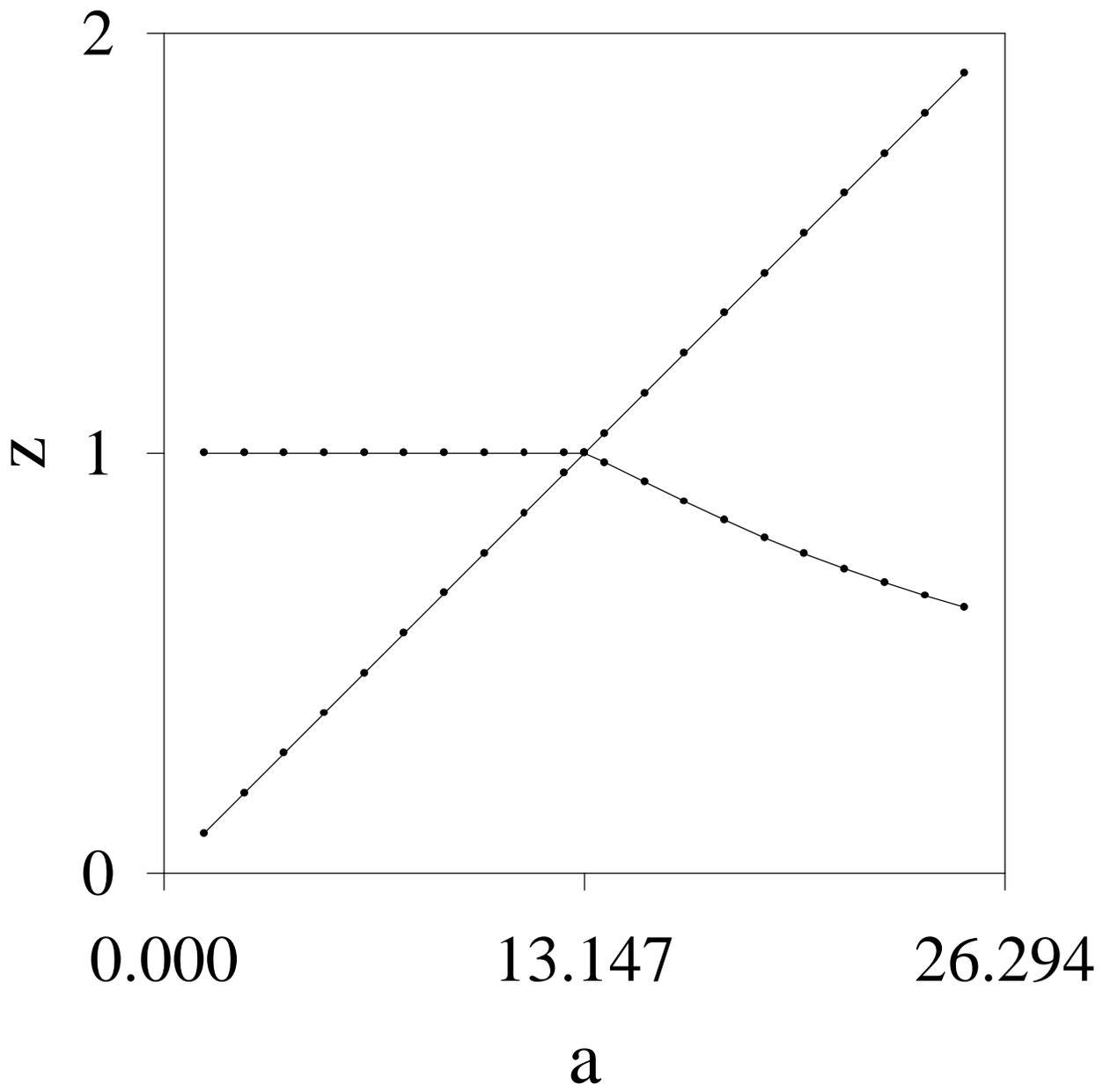
\end{document}